%% file: Aluie_arXiv.tex
\documentclass[]{siamart0516}

\input{ex_shared}

\ifpdf
\hypersetup{
  pdftitle={\TheTitle},
  pdfauthor={\TheAuthors}
}
\fi

\begin{document}

\maketitle

% REQUIRED
\begin{abstract}
We generalize the definition of convolution of vectors and tensors on the 2-sphere, and prove that it commutes with differential operators. Moreover, vectors and tensors that are normal/tangent to the spherical surface remain so after the convolution. These properties make the new filtering operation particularly useful to analyzing and modeling nonlinear dynamics in spherical systems, such as in geophysics, astrophysics, and in inertial confinement fusion applications. An essential tool we use is the theory of scalar, vector, and tensor spherical harmonics. We then show that our generalized filtering operation is equivalent to the (traditional) convolution of scalar fields of the Helmholtz decomposition of vectors and tensors.
\end{abstract}

\vspace{.5cm}
\section*{Preamble}
This work has been accepted for publication in {\it GEM - International Journal on Geomathematics}, which is fitting since
the results herein rely in a crucial manner on results by {\it GEM}'s Editor-in-Chief, professor Willi Freeden, and his collaborators.

The paper was under prior consideration in the journal {\it Nonlinearity}. However, after more than one year in the review process and despite the positive review from the referee, the editors deemed the manuscript not a good fit for that journal. To shed more light on those circumstances, the correspondence from the editors of {\it Nonlinearity} along with the referee review can be found here: \url{http://www.complexflowgroup.com/links/Correspondence.pdf}. 

\section{Introduction\lb{sec:Introduction}}
In this paper, we generalize the traditional convolution on a spherical surface to ensure that it commutes with differential operators on the 2-sphere, $\cS$. This work constitutes the mathematical foundation for recent studies of energy pathways across scales in global oceanic flows \cite{Aluieetal18,Sadeketal17}.
To illustrate the type of problems motivating this work, consider as an example the Navier-Stokes equation over $\mathbb{R}^3$,
\vspace{.5cm}
\be
\partial_t \bu + \grad\bdot(\bu\otimes\bu) = -\grad{P} + \nu\Delta\bu, \hspace{2cm} \grad\bdot\bu=0,
\lb{eq:NS}\ee
supplemented with appropriate initial conditions. Here, $\bu(\bx;t)$ is a velocity vector field solution, $\otimes$ is a dyadic tensor product, $\nu$ is a constant representing viscosity, and $P(\bx;t)$ is a Lagrange multiplier \cite{DoeringGibbon95} satisfying a Poisson equation and is solely a function of $\bu(\bx;t)$. A standard approach in fluid dynamics is to derive a filtered version of equation (\ref{eq:NS}):
\vspace{.5cm}
\be
\partial_t \widetilde\bu_\ell + \grad\bdot(\widetilde{\bu}_\ell\otimes\widetilde{\bu}_\ell) = -\grad{\widetilde{P}_\ell} + \nu\Delta\widetilde{\bu}_\ell -\grad\bdot(\widetilde{\bu\otimes\bu}_\ell - \widetilde{\bu}_\ell\otimes\widetilde{\bu}_\ell),\hspace{.75cm} \grad\bdot\widetilde\bu_\ell=0,
\lb{eq:NS_fltr}\ee
where
\be\widetilde f_\ell =  \int_{\mathbb{R}^3} \,d^3\br\, G(\br-\bx; \ell) \,f(\br)
\lb{eq:FilteringEuclid}\ee 
is a filtered field obtained by convolving an integrable scalar function $f(\bx)$ with a real-valued kernel $G(\ell)$ that has characteristic spatial width $\ell$. Operation (\ref{eq:FilteringEuclid}) can be applied to vector and higher order tensor fields by operating on each of the Cartesian components separately.
Eq. (\ref{eq:NS_fltr}) is exact and follows from applying filtering operation (\ref{eq:FilteringEuclid}) to eq. (\ref{eq:NS}), and using \textbf{a key property that operation (\ref{eq:FilteringEuclid}), when applied to a tensor of any rank, commutes with spatial derivatives in Euclidean space}\footnote{That operation (\ref{eq:FilteringEuclid}) also commutes with time derivatives is trivial.}. For example, we have  $\widetilde{\grad P}_\ell= \grad \widetilde{P}_\ell$, $\widetilde{\grad\bdot\bu}_\ell=\grad\bdot\widetilde\bu_\ell$, $\widetilde{\Delta\bu}_\ell = \Delta\widetilde{\bu}_\ell$, and $\widetilde{\grad\bdot(\bu\otimes\bu)}_\ell = \grad\bdot(\widetilde{\bu\otimes\bu}_\ell)$. This allows filtered eq. (\ref{eq:NS_fltr}) to resemble the original Navier-Stokes dynamics, thereby allowing a straightforward interpretation of the various terms. The only difference is the additional term, $\grad\bdot(\widetilde{\bu\otimes\bu}_\ell - \widetilde{\bu}_\ell\otimes\widetilde{\bu}_\ell)$, arising from the nonlinearity and accounting for interactions with spatial scales that are filtered out.  
 A great deal of effort then goes into developing a physical understanding of such a term and deriving approximations to it that can be used in simulations at a lower spatial resolution without having to explicitly capture all scales that exist in natural or laboratory flows. The aforementioned goals comprise the primary focus of Large Eddy Simulation (LES) modeling \cite{Leonard74,Germano92,GalperinOrszag93,MeneveauKatz00,Pope04,Pitsch06,Sagaut06,Garnieretal09}, which is a wide field of research in fluid dynamics and turbulence.
 
The filtering methodology has also been developed and utilized extensively as both a mathematical analysis tool of partial differential equations (PDEs) (e.g. \cite{Evans98}) and a physics diagnostic tool in turbulence research, which has led to physical insight and the derivation of mathematically exact estimates 
(e.g. \cite{Eyink95a,Chenetal03,Eyink05,GeurtsHolm06,LinshizTiti07,EyinkAluie09,AluieEyink10,Aluie11,AluieKurien11,DascaliucGrujic11,Aluie13,Riveraetal14,Aluie17}). 
Such an approach goes beyond a mere scale decomposition of a signal to the multiscale analysis of its \emph{dynamics} through the governing PDE. Kernel $G(\ell)$ used in operation (\ref{eq:FilteringEuclid}) can belong to a wide class of functions, including Schwartz functions, radial basis functions, and scaling functions and wavelets commonly used in fields such as approximation theory, signal processing, and texture analysis.

\subsection{On the Sphere}
The generalized convolution we present here would allow for the extension of the above approach to analyzing PDEs on the sphere. An obvious area of application is in geophysics and climate, where the ever-increasing availability and accessibility of global Earth data makes such mathematical frameworks for describing multiscale processes all the more pertinent. Another area of application is astrophysics, where large surveys of the sky using land- and space-based telescopes are yielding massive amounts of data to be queried. Yet, a third and perhaps less known area of application that can benefit from extending such an approach to spherical manifolds is inertial confinement fusion (ICF) \cite{BettiHurricane16}. In ICF, a spherical cryogenic target which comprises mainly of a hydrogen ice-shell enveloping hydrogen gas, is imploded with powerful lasers in a spherically symmetric fashion in an effort to initiate a nuclear reaction, similar to that occurring in the Sun, with the potential of yielding virtually limitless amounts of clean energy. Other potential areas of application include solar physics, planetary physics, and magnetospheric processes.

The literature in applied spherical harmonic analysis is vast and includes topics such as spherical convolutions and wavelets (e.g. \cite{SchroderSweldens95,AntoineVandergheynst99,Bayeretal01,Antoineetal02,MayerMaier06,FreedenGerhards10}), the computation of spherical harmonic coefficients (e.g. \cite{DriscollHealy94, Mohlenkamp99,Healyetal03}) and the computation of convolutions (e.g. \cite{WandeltGorski01, BohmePotts03, KeinerPrestin06}), and spatially localized spectral modeling of data (e.g.\cite{Schmidtetal06,WieczorekSimons05}). Many studies have focused on studying and applying specific kernels, such as the Abel-Poisson in \cite{Fleischmannetal10}, or classes of kernels such as correlation functions, splines, and wavelets (e.g.\cite{Alfeldetal96,BernsteinEbert10}). Other works have addressed the Helmholtz decomposition of vectors and tensors on $\cS$ using a variety of tools such as scaling functions, wavelets, splines, or spherical harmonics (e.g. \cite{Wahba82,FreedenGervens93,FenglerFreeden05,FuselierWright09,FreedenSchreiner09,FreedenGerhards10,FuselierWright16}).  The motivation has often stemmed from application areas such as the image analysis of faces, the analysis and modeling of geophysical data from satellites, or of astrophysical telescope data, to name a few. 

Significant developments have been driven by approximation theory (e.g. \cite{Gutzmer96,FasshauerSchumaker98,Freedenetal98,YershovaLaValle04,Hielscheretal10,BerkelMichel10,FilbirPotts10,Gneiting13}), where localized basis functions are used for interpolating scattered data on $\cS$. 
Our work here has substantial connections to the radial basis functions (RBFs) literature within approximation theory. There, the goal is to solve for $f(\bx_i)$ the linear convolution,
$$
\widetilde f_\ell(\bx) = G(\ell) * f(\bx),
$$
given input data $\widetilde f_\ell(\bx_i)$ over scattered spatial locations $\bx_i$, where $i=1,...,N$. The resulting `interpolant' $\widetilde f_\ell(\bx)$ can then be evaluated over any location $\bx$ in the domain.
The convolution kernel $G$ is an RBF, and parameter $\ell$ is not thought of as a length-scale of a filter but as a 'shape parameter' that can be adjusted to yield the highest accuracy of interpolation while maintaining stability. Therefore, solving the interpolation problem is equivalent to a de-convolution, where much effort has been devoted to address the ill-conditioning of the problem (e.g. \cite{Mairhuber56,Hardy71,FornbergWright04, FornbergPiret07,FornbergFlyer15b}). The interpolation problem applied to vectorial data, as was done for example in 
\cite{Fehlingeretal07}, has some relevance to our work here. The goal in 
\cite{Fehlingeretal07} was to reconstruct an approximation of the local sea surface height $h$, which is a scalar field, from velocity field measurements, $\bv$, by solving the surface curl gradient equation
$$
\bv(\bx)=\hat\Be_r\btimes\grad^*\widetilde{h}_\ell(\bx),
$$
where $\hat\Be_r$ is the radial unit vector normal to $\cS$ and $\grad^*$ is the gradient tangent to $\cS$. The velocity $\bv(\bx_i)$ is known over points $\bx_i$, $i=1,..., N$. In \cite{Fehlingeretal07}, a regularized Green's function with respect to the Beltrami operator which can be thought of as a special choice for kernel $G(\ell)$, was used to prove that solution $\widetilde{h}_\ell$ of the above equation, converged uniformly to the exact field $h$ in the limit $\ell\to 0$. However \cite{Fehlingeretal07} did not discuss the relation between $\widetilde h_\ell$ and the convolution of vector field $\hat\Be_r\btimes\grad^*h(\bx)$ for \emph{any fixed $\ell > 0$}. One of our goals here is determining the proper convolution of a vector field such as $\hat\Be_r\btimes\grad^*h(\bx)$ and its relation to the convolution of scalar field $h(\bx)$.

While our work builds upon or overlaps with some of the aforementioned studies, \textbf{a key difference is our interest in ensuring that the filtering operation commutes with differential operators on $\cS$}. To the best of our knowledge, this issue has not been addressed in the published literature. 
Perhaps the work with goals closest to ours is \cite{Narcowichetal07,Fuselieretal09}, where the authors developed divergence-free and curl-free RBFs on general curved surfaces, including on the sphere. Those studies recognized that divergence-free RBFs in $\cR^3$ \cite{NarcowichWard94}, when projected onto an embedded surface, do not remain divergence-free. The reason being that while convolution operation (\ref{eq:FilteringEuclid}) and spatial derivatives commute in Euclidean space such as $\cR^3$, they no longer commute when restricted onto an embedded curved manifold. However, our goal here extends beyond imposing the divergence-free or curl-free property to having, for example, $\grad\bdot\widetilde{\bu}_\ell=\widetilde{f}_\ell(\bx) = \Delta\widetilde\psi_\ell$ if the underlying vector field satisfies $\grad\bdot\bu=f(\bx)= \Delta\psi$. In this respect, our work can be viewed as a generalization of that by \cite{Narcowichetal07,Fuselieretal09}, although we shall follow an approach that is completely different from theirs. We shall also tackle differential operators other than the divergence and curl, and our results extend to rank-2 tensors in addition to vectors.
 
Our results below rely heavily on work by Willi Freeden and co-authors (e.g. \cite{Freedenetal98,FreedenSchreiner09}), especially their generalization of the Funk-Hecke theorem to vectors and tensors. Following their work, we also use a coordinate-free approach in our proofs to avoid pole singularities inherent to any spherical coordinate system. 

This paper is organized as follows. In sections \ref{sec:Preliminaries} we present preliminaries and a brief recap of the theory of spherical harmonics. We present our generalization of the convolution to vectors in section \ref{sec:Vectors}, and to tensors in section \ref{sec:Tensors}. In section \ref{sec:SphericalFiltering}, we define a filtering operation based on the generalized convolution and show that it satisfies certain desired properties. Section \ref{sec:FilteringDerivativesCommute} proves that our new filtering operation and differential operators on the sphere commute. Section \ref{sec:FilteringWOsfts} shows that filtering vectors and tensors is equivalent to filtering the scalar fields resultant from a Helmholtz decomposition. We conclude the paper with section \ref{sec:Conclusion}.

\section{Preliminaries and Notation\lb{sec:Preliminaries}}
We shall work on the surface of a sphere of radius $r$, denoted by $\cS$. With the canonical Cartesian basis in $\cR^3$, $\hat{\Be}_i$ for $i=1,2,3$, and vector $\br=(r_1,r_2,r_3)^T=\sum_{i=1}^3 r_i\hat{\Be}_i\in\cR^3$  of Euclidean length $|\br|=\sqrt{r_1^2+r_2^2+r_3^2}$, this space can be defined as $\cS=\left\{\br\in\cR^3~\big|~|\br|=r \right\}$.
 Let $R \in$ SO(3) denote a rotation on $\cS$. The transpose of $R$ is its inverse, such that $RR^{T}=I$ yields the identity matrix, $I$. Its determinant is $\mbox{det}|R| = +1$. The rotation of a scalar function, $f(\br)$, on $\cS$ is defined as
\be
{\mathcal R}_R ~f(\br) \coloneqq  f(R^{-1}\br).
\ee
The surface area element on the sphere, $dS(\br)$, is the unique measure invariant under the SO(3) group. Its integral is $\int_{\cS} dS(\br) = 4\pi r^2$.

We denote the space of $p$-integrable scalar, vector, and tensor real-valued fields over domain $\Omega$ by $L^p(\Omega)$, {\bf l}$^p(\Omega)$, and $\bL^p(\Omega)$, respectively. Similarly, we denote by the space of scalar, vector, and tensor real-valued fields that are $k$-times continuously differentiable over $\Omega$ by $C^{(k)}(\Omega)$, ${\bf c}^{(k)}(\Omega)$, and ${\bf C}^{(k)}(\Omega)$, respectively ($k=0$ represents the space of continuous fields). 

The inner product over Hilbert space $L^2(\cS)$ is
\be \langle f,g \rangle \coloneqq  {1\over 4\pi r^2} \int_{\cS} ~dS(\br) \,f(\br)\,g(\br), \hspace{1cm} f,g\in L^2(\cS).
\lb{eq:def_innerproduct}\ee
A spatial average over $\cS$ is denoted by
\be \langle f\rangle \coloneqq   \langle f,1 \rangle = {1\over 4\pi r^2} \int_{\cS} ~dS(\br) \,f(\br).
\lb{eq:def_average}\ee
A useful adjoint property is 
\be 
\int_{\cS} ~dS(\br) ~f(R^{-1}\br) ~g(\br) = \int_{\cS} dS(\br) ~f(\br) ~g(R\br), 
\ee
which follows from the invariance of the measure under rotations,\\
$dS(R^{-1}\br) = \mbox{det}|R^{-1}| \,dS(\br)$
(e.g. \cite{FreedenSchreiner09,Michel13}). 

Geodesic distances (or scalar products of position vectors) on the sphere are invariant to rotation: for $\br_1,\br_2\in\cS$, we have 
$\mbox{dist}(R\,\br_1,R\,\br_2)=\mbox{dist}(\br_1,\br_2)$, where 
$$ \mbox{dist}(\br_1,\br_2) = r \cos^{-1}\left[ \frac{\br_1\bdot\br_2}{r^2}\right].
$$

A function $G_\bx$ that is only a function of geodesic distance from $\bx\in\cS$ is called an $\bx$-zonal function on $\cS$ (e.g. \cite{Michel13}),
\be G_\bx:\cS\to\cR \hspace{1cm} G_\bx(\br) \coloneqq G\left(\frac{\bx\bdot\br}{r^2}\right),
\ee
where $G:[-1,1]\to \cR$. 

For a zonal function, $G_\bx$ on $\cS$, such that $G \in$ $L^p[-1,1]$, and scalar field, $f\in$ $L^q(\cS)$ with $p,q\in(1,\infty)$ and $1/p+1/q=1$, a convolution on $\cS$ is defined as
\be G_\bx * f \coloneqq \int_\cS ~dS(\br)\, G\left(\frac{\bx\bdot\br}{r^2}\right) \,f(\br)~.
\lb{eq:ConvolutionScalar}\ee
Moreover, the zonal function is said to be normalized iff 
\be G_\bx * 1 = 1~~\Longleftrightarrow \int_{-1}^1dt\,G(t)=(2\pi r^2)^{-1}.
\lb{eq:normalizedDef}\ee

\subsection{Differential operators on the sphere}
In what follows, we use  spherical coordinates $\br=(r,\lambda,\phi)$, where $\lambda \in[-\pi,\pi]$ is longitude and $\phi \in[-\pi/2,\pi/2]$ is latitude, with the poles at $\phi=\pm\pi/2$. 
The gradient of scalar field $g\in C^{(1)}(\cR^3)$ in spherical coordinates is
\be \grad g= \hat{\Be}_r \partial_r g
+ \hat{\Be}_\lambda {1\over r \cos\phi} \partial_\lambda g
+ \hat{\Be}_\phi  {1\over r} \partial_\phi g,
\ee
where $\hat{\Be}_r$, $\hat{\Be}_\lambda$, and $\hat{\Be}_\phi$ are the canonical orthonormal spherical coordinates basis in $\cR^3$  such that $\hat{\Be}_r\btimes \hat{\Be}_\lambda = \hat{\Be}_\phi$.
The normalized gradient tangent to $\cS$ for scalar field $f\in C^{(1)}(\cS)$  is
\be \grad^* f\coloneqq
\hat{\Be}_\lambda {1\over  \cos\phi} \partial_\lambda f
+ \hat{\Be}_\phi  \partial_\phi f.
\lb{eq:TangentGradient}\ee
For vector field $\bv\in {\bf c}^{(1)}(\cR^3)$, the divergence in spherical coordinates is
\be \grad\bdot\bv 
= {1\over r^2}\partial_r \left( r^2 v_r \right) 
+ {1\over r\cos\phi} \partial_\lambda v_\lambda + {1\over r\cos\phi} \partial_\phi\left( v_\phi \cos\phi\right),
\ee
where $v_r = \bv\bdot \hat{\Be}_r$, $v_\lambda = \bv\bdot \hat{\Be}_\lambda$, and $v_\phi = \bv\bdot \hat{\Be}_\phi$.
The normalized divergence tangent to $\cS$, for $\bu\in {\bf c}^{(1)}(\cS)$,  is
\be \grad^*\bdot\bu  
\coloneqq {1\over \cos\phi} \partial_\lambda u_\lambda + {1\over \cos\phi} \partial_\phi\left( u_\phi \cos\phi\right),
\lb{eq:TangentDivergence}\ee
The curl in spherical coordinates is
\begin{eqnarray}
\grad\btimes\bv &=& \frac{1}{r^2 \cos\phi}\mbox{det}\Bigg|
\begin{array}{ccc}
\hat{\Be}_\lambda r\cos\phi ~&~ \hat{\Be}_\phi r ~&~ \hat{\Be}_r\\
\partial_\lambda ~&~ \partial_\phi ~&~ \partial_r\\
v_\lambda r\cos\phi ~&~ v_\phi r ~&~ v_r
\end{array}
\Bigg|,
\lb{eq:CurlSpherical3Ddefinition}\end{eqnarray}
The normalized curl normal to $\cS$ is
\begin{eqnarray}
\grad^*\btimes\bu 
\coloneqq  \hat{\Be}_r\frac{1}{\cos\phi}\left[\partial_\lambda u_\phi - \partial_\phi( u_\lambda \cos\phi)\right].
\lb{eq:TangentCurl}\end{eqnarray}
Note that definitions (\ref{eq:TangentDivergence}) and (\ref{eq:TangentCurl}) are independent of the radial component of a vector field. 

Borrowing notation from Freeden and Schreiner \cite{FreedenSchreiner09}, we also introduce two related curl operators which we shall use in this paper. The first acts on scalars and yields a vector, whereas the second acts on vectors and yields a scalar, analogous to the gradient and divergence operators, respectively:
\begin{eqnarray}
\bL^* g &\coloneqq&  \hat{\Be}_r\btimes \grad^* g = -r\grad\btimes(\hat{\Be}_r \, g)\lb{eq:Lstar}\\
\bL^* \bdot \bu &\coloneqq&  (\grad^*\btimes\bu)\bdot\hat{\Be}_r\lb{eq:LstarDot}
\end{eqnarray}
Note that $\grad^*\btimes\bu = (\bL^* \bdot \bu)\hat{\Be}_r$ is normal to $\cS$, whereas $\bL^* g$ is tangent to $\cS$.

The curl can be rewritten in coordinate-free form, which will be useful in some of our proofs below:
\begin{eqnarray}
\grad\btimes\bv = \frac{1}{r}\grad^*\btimes\bv + \frac{1}{r}(\grad^*v_r)\btimes\hat{\Be}_r - \frac{1}{r}\partial_r\left[r\,\bv\btimes\hat{\Be}_r\right].
\lb{eq:Curl3DCoordFree}
\end{eqnarray}

The Laplacian in spherical coordinates, for $g\in C^{(2)}(\cR^3)$, is
\be \Delta g = {1\over r^2}\left[\partial_r\left(r^2\partial_r g\right)
+{1\over \cos\phi}\partial_\phi\left(\cos\phi~\partial_\phi g\right)
+{1\over \cos^2\phi}\partial^2_\lambda g			\right]
\ee
The normalized Laplacian on $\cS$, often called the Beltrami operator, is
\be \Delta^*f=
{1\over \cos\phi}\partial_\phi\left(\cos\phi~\partial_\phi f\right)
+{1\over \cos^2\phi}\partial^2_\lambda f,
\ee
 where $f\in C^{(2)}(\cS)$. It can be rewritten as
\be
\Delta^*f = \grad^*\bdot\grad^* f =\bL^*\bdot\bL^* f.
\ee

\subsection{Spherical Harmonics\lb{sec:SphericalHarmonics}}
In our proofs, we make extensive use of spherical harmonics which we shall recap briefly in this section. More thorough expositions may be found in many standard references in mathematics and physics (e.g. \cite{Edmonds57,Groemer96,BiedenharnLouck09,AtkinsonHan12}).

\subsubsection{Spherical harmonics of scalar fields}
Any function $f(\br)\in$ $L^2(\cS)$  can be decomposed into spherical harmonics
\be f(\br) = \sum_{n,j} \hat{f}_{n,j}(r) ~ Y_{n,j}(\lambda,\phi),
\ee
where $\br=(r,\lambda,\phi)$ and $Y_{n,j}(\br/r)$ is a real-valued spherical harmonic of degree $n$ and order $j$, following \cite{FreedenSchreiner09}. It is a function of position $(\lambda,\phi)$ on  $\cS$ and is independent of radius $r$. The degree $n\in \cN$ is a non-negative integer and $j$ is any integer between $-n,\dots, -1, 0, 1, \dots, n$. The set $\{Y_{n,j}\},\,n=0,1,\dots,\, j=-n,\dots,n$ forms an orthonormal system in the space of square-integrable functions on $\cS$:
\be \langle Y_{n,j}\, Y_{m,k}\rangle = {1\over 4\pi r^2} \int_{\cS} ~dS(\br)~Y_{n,j}\left(\frac{\br}{r}\right) ~Y_{m,k}\left(\frac{\br}{r}\right) = \delta_{nm}\delta_{jk}.
\ee
$Y_{n,j}$ are also eigenfunctions of the Beltrami operator:
\be \Delta^* Y_{n,j} = -n(n+1)Y_{n,j} 
\ee

The spherical Fourier transform (SFT) of $f(\br)$ on $\cS$ is:
\be {\mathcal {SFT}}\left\{f \right\} \coloneqq\hat{f}_{n,j}(r) = \frac{1}{4\pi r^2}\int_{\cS} dS(\br) \,f(\br) ~Y_{n,j}\left(\frac{\br}{r}\right).
\ee

The SFT satisfies Parseval's identity:
\be \| f \|^2_{L^2(\cS)} = \sum_{n=0}^\infty\sum_{j=-n}^n |\hat{f}(n,j)|^2,
\ee
where the $L^p$ norm over $\cS$ is 
\be \|\dots\|_{L^p(\cS)} \coloneqq \left\langle\left(\dots\right)^p\right\rangle^{1/p}
\ee

The Legendre polynomial (Rodriguez Formula), 
\begin{eqnarray} 
P_n(t) 
&=& \frac{1}{2^n\,n!}\left(\frac{d}{dt}\right)^n (t^2-1)^n,\lb{eq:LegendrePolynomial}
\end{eqnarray}
is a polynomial of degree $n$, where $t=\bx\bdot\br/r^2\in[-1,1]$ for $\bx$, $\br \in \cS$. Legendre polynomials are orthogonal, $\int_{-1}^{1}dt ~ P_n(t) \, P_m(t) = \delta_{nm}\,2/(2n+1)$, and satisfy $P_n(1)=1$. They are related to spherical harmonics through the {\it addition theorem}:
\be P_n\left(\frac{\bx\bdot\br}{r^2}\right) = \frac{1}{2n+1}\sum_{j=-n}^{n} Y_{n,j}\left(\frac{\bx}{r}\right)Y_{n,j}\left(\frac{\br}{r}\right),
\ee
for $\bx$, $\br \in \cS$. The {\it Legendre transform} of a function $G\in L^1[-1,1]$ is defined as
\begin{eqnarray} 
{\mathcal {LT}}\left\{G \right\} \coloneqq \widehat{G}(n) &=& 2\pi r^2 \int^1_{-1} dt~G(t) \, P_n(t)\lb{eq:LegendreTransform}\\
&=& \int_\cS dS(\br)~G\left(\frac{\bx\bdot\br}{r^2}\right) P_n\left(\frac{\bx\bdot\br}{r^2}\right).\nonumber
\end{eqnarray} 
 Its inverse is
\be {\mathcal {LT}^{-1}}\left\{\widehat{G}(n) \right\} \coloneqq G(t) = \sum_{n=0}^\infty \widehat{G}(n) \frac{2n+1}{4\pi r^2}P_n(t).
\lb{eq:InvLegendreTransform}\ee 

A cornerstone result in harmonic analysis on $\cS$ is the spherical convolution theorem  due to Funk \cite{Funk1916} and Hecke \cite{Hecke1918}. For $G \in L^1[-1,1]$,
\be \int_{\cS}dS(\br) \,G\left(\frac{\bx\bdot\br}{r^2}\right) P_n\left(\frac{\by\bdot\br}{r^2}\right)= \widehat{G}(n)\,P_n\left(\frac{\bx\bdot\by}{r^2}\right),
\ee
for $\bx$, $\by \in \cS$. An important corollary of the Funk-Hecke theorem relates convolutions of functions to the product of their Fourier coefficients:
\be {\mathcal {SFT}} \left\{G_\bx*f \right\}= \widehat{G}(n)\hat{f}_{n,j}(r),
\ee
where $G \in L^q[-1,1]$ and $f\in L^p(\cS)$ with $p,q\in(1,\infty)$ and $1/p +1/q = 1$ (e.g. \cite{Michel13}).

\subsubsection{Spherical harmonics of vector fields}
The orthonormal vector spherical harmonic basis on $\cS$ is:
\begin{eqnarray}
\bY_{n,j} &\equiv& \hat{\Be}_r \,Y_{n,j}\lb{eq:VectorBasis1}\\
\bPsi_{n,j} &\equiv&  {1\over \sqrt{n(n+1)}} \grad^* Y_{n,j}\lb{eq:VectorBasis2}\\
\bPhi_{n,j} &\equiv& {1\over \sqrt{n(n+1)}} \bL^* Y_{n,j} \lb{eq:VectorBasis3}
\end{eqnarray}
$\bY_{n,j}$ is normal to the spherical surface while $\bPsi_{n,j}$ and $\bPhi_{n,j}$ are tangent to it. The gradient and curl operators, $\grad^*$ and $\bL^*$, are defined in eqs. (\ref{eq:TangentGradient}),(\ref{eq:Lstar}), respectively. $\bPsi_{n,j}$ is irrotational, $\bL^*\bdot\bPsi_{n,j}=0$, while $\bPhi_{n,j}$ is solenoidal (or toroidal), $\grad^*\bdot\bPhi_{n,j}=0$.

\section{Convolution of Vector Fields\lb{sec:Vectors}}
A straightforward application of the convolution defined in eq. (\ref{eq:ConvolutionScalar}) to a vector is equivalent to convolving each of its Cartesian components as scalars:
\begin{eqnarray} 
G_\bx*\bu = (G_\bx*u_1) \hat{\Be}_1 + (G_\bx*u_2) \hat{\Be}_2 + (G_\bx*u_3) \hat{\Be}_3,
\lb{eq:TraditionalConvVectors}\end{eqnarray} 
where $\hat{\Be}_i$, for $i=1,2,3$, are the canonical Cartesian orthonormal basis vectors in $\cR^3$.
However, such a convolution applied to vectors and tensors does not commute with differential operators.
As we mentioned in the Introduction, this is important to derive tractable coarse-grained PDEs which govern the evolution of scales. For instance, if the original vector field is divergence-free, we would want the coarse-grained field to be divergence-free as well. However, coarse-graining using the convolution in eq. (\ref{eq:ConvolutionScalar}) fails these commutation requirements:
$$G_\bx*(\grad\btimes\bu) \ne \grad\btimes(G_\bx*\bu), 
\hspace{.4cm} G_\bx*(\grad f) \ne \grad (G_\bx*f),
\hspace{.4cm} G_\bx*(\grad \bdot\bu) \ne \grad \bdot(G_\bx*\bu).
$$
A related issue concerns purely radial  and purely tangent vector fields, $f\hat{\Be}_r$ and $\bu_t$, respectively, which do not remain radial  or tangent when convolved according to (\ref{eq:TraditionalConvVectors}):
$$G_\bx*(f \, \hat{\Be}_r) \ne (G_\bx*f) \, \hat{\Be}_r, 
\hspace{.4cm} G_\bx*(\bu \btimes \hat{\Be}_r) \ne (G_\bx*\bu) \btimes \hat{\Be}_r.
$$
This makes deriving coarse-grained PDEs and the analysis of scale dynamics on $\cS$ intractable.
The problem, as we shall see later, stems from that whereas the scalar spherical harmonic basis functions, $Y_{n,j}$, are eigenfunctions of the convolution operation, $G*(\cdot)$, the same is not true for \emph{vector} spherical harmonic basis functions in eqs. (\ref{eq:VectorBasis1})-(\ref{eq:VectorBasis3}). 

To this end, in addition to the orthonormal vector basis (\ref{eq:VectorBasis1})-(\ref{eq:VectorBasis3}) above,  we will need to use another orthonormal basis due to Edmonds \cite{Edmonds57} (see also \cite{FreedenSchreiner09}):
\begin{eqnarray}
\bK^{(1)}_{n,j} &=& \frac{1}{\sqrt{(n+1)(2n+1)}} \left[\left(\left(2n+1\right)\br - r^2\grad\right) r^n Y_{n,j}\right]_{r=1}\lb{eq:EdmondsBasis1}\\
\bK^{(2)}_{n,j} &=& \frac{1}{\sqrt{n(2n+1)}} \left[\grad (r^n Y_{n,j})\right]_{r=1}\lb{eq:EdmondsBasis2}\\
\bK^{(3)}_{n,j} &=& \frac{1}{\sqrt{n(n+1)}} \left[\br\btimes\grad(r^n Y_{n,j})\right]_{r=1}\lb{eq:EdmondsBasis3}
\end{eqnarray}
Unlike  basis (\ref{eq:VectorBasis1})-(\ref{eq:VectorBasis3}) which consists of radial and tangential fields, this basis, which we shall call the {\it Edmonds basis}, does not permit this simple geometric interpretation. However, the advantage of the Edmonds basis functions lies in that the Cartesian vector components of each of $\bK^{(1)}_{n,j}$, $\bK^{(2)}_{n,j}$, and $\bK^{(3)}_{n,j}$ is a scalar spherical harmonic of degree $n+1$, $n-1$, and $n$, respectively \cite{Freedenetal98}. They arise from restricting to the unit sphere homogeneous harmonic \emph{vector} polynomials (terms in brackets in eqs. (\ref{eq:EdmondsBasis1})-(\ref{eq:EdmondsBasis3})) of degree $n+1$, $n-1$, and $n$, respectively. We remind the reader that a vector $\bu$ in $\cR^3$ is called a homogeneous harmonic \emph{vector} polynomial of degree $n$ if each of its Cartesian components, $u_1=\bu\bdot\hat{\Be}_1$, $u_2=\bu\bdot\hat{\Be}_2$, $u_3=\bu\bdot\hat{\Be}_3$ is a scalar homogeneous harmonic polynomial of degree $n$. The basis $\bY_{n,j} $, $\bPsi_{n,j}$, and $\bPhi_{n,j}$ are \emph{not} a result of restricting homogeneous harmonic vector polynomials of degree $n$ to a spherical surface. This is why $\bY_{n,j} $, $\bPsi_{n,j}$, and $\bPhi_{n,j}$ are not eigenfunctions of the convolution operation (\ref{eq:ConvolutionScalar}), in contrast to $\bK^{(1)}_{n,j}$, $\bK^{(2)}_{n,j}$, and $\bK^{(3)}_{n,j}$ (see \cite{FreedenSchreiner09} for details):
\begin{eqnarray}
G_\bx* \bK^{(i)}_{n,j} &=& \sum_{m=1}^3 G_\bx* \left(\bK^{(i)}_{n,j}\right)_m\,\hat{\Be}_m 
= \widehat{G}\left(\mbox{deg}\left(i \right)\right) \,\bK^{(i)}_{n,j}		\hspace{1cm} i=1,2,3
\lb{eq:ConvEdmondsBasis}\end{eqnarray}
where $\left(\bK^{(i)}_{n,j}\right)_m = \bK^{(i)}_{n,j}\,\bdot\,\hat{\Be}_m$, is the $m$-th Cartesian component for $m=1,2,3$, $\widehat{G}$ is the Legendre transform, as defined in eq. (\ref{eq:LegendreTransform}), and
\begin{eqnarray}
\mbox{deg}\left(i \right) 
&=&\Bigg\{
\begin{array}{c c}
n+1 & ~~ i=1\\
n-1  & ~~ i=2\\
n     & ~~ i=3\\
\end{array}
\end{eqnarray}

The Edmonds basis is related to basis (\ref{eq:VectorBasis1})-(\ref{eq:VectorBasis3}) through what we shall dub the Edmonds transformation: 
\be
\left(
\begin{array}{c}
\bK^{(1)}_{n,j}\\
\bK^{(2)}_{n,j}\\
\bK^{(3)}_{n,j}\\
\end{array}
\right)
=
[S]\left(
\begin{array}{c}
\bY_{n,j}\\
\bPsi_{n,j}\\
\bPhi_{n,j}\\
\end{array}
\right)
\ee
where
\be [S]=
\frac{1}{\sqrt{2n+1}}\left(
\begin{array}{ccc}
\sqrt{n+1} & -\sqrt{n} & 0 \\
 \sqrt{n} & \sqrt{n+1} & 0  \\
 0 & 0 & \sqrt{2n+1}  \\
\end{array}
\right)
\ee
is a unitary matrix with its inverse $[S^{-1}]=[S^{\mbox{{\small T}}}]$.

Any vector $\bu\in$ {\bf l}$^2(\cS)$ has the following Fourier representation:
\begin{eqnarray}
\bu(\br) &=& \sum_{n=0}^\infty \sum_{j=-n}^n \hat{u}^Y_{n,j}(r)  \bY_{n,j} + \hat{u}^\psi_{n,j}(r)  \bPsi_{n,j}  +  \hat{u}^\phi_{n,j}(r)  \bPhi_{n,j} \lb{eq:VectorSFT_1}\\
&=& \sum_{n=0}^\infty \sum_{j=-n}^n \hat{u}^{(1)}_{n,j}(r)  \bK^{(1)}_{n,j} + \hat{u}^{(2)}_{n,j}(r)  \bK^{(2)}_{n,j} +  \hat{u}^{(3)}_{n,j}(r)   \bK^{(3)}_{n,j} \lb{eq:VectorSFT_2}
\end{eqnarray}
where the Fourier coefficients are obtained from the following vector Spherical Fourier Transform (vSFT):
\begin{eqnarray}
\hat{u}^Y_{n,j}(r) &=&\langle\bu\bdot\bY_{n,j}\rangle, \hspace{.5cm}
\hat{u}^\psi_{n,j}(r) =\langle\bu\bdot\bPsi_{n,j}\rangle, \hspace{.5cm}
\hat{u}^\phi_{n,j}(r) =\langle\bu\bdot\bPhi_{n,j}\rangle\lb{eq:vSFT}\\
\hat{u}^{(i)}_{n,j}(r) &=&\langle\bu\bdot \bK^{(i)}_{n,j}\rangle, \hspace{.5cm} i=1,2,3\lb{eq:EdmondsvSFT}
\end{eqnarray}
In light of this, any vector field may be decomposed into three parts:
\be \bu(\br) = \bu^{(1)}(\br)+\bu^{(2)}(\br)+\bu^{(3)}(\br)
\lb{eq:VectorDecompostion_1}\ee
with
\be
\bu^{(i)}(\br) \coloneqq \sum_{n=0}^\infty \sum_{j=-n}^n \hat{u}^{(i)}_{n,j}(r)  \bK^{(i)}_{n,j}, \hspace{1cm} i=1,2,3
\lb{eq:EdmondsVectors}\ee
Note that the decomposition in eq.(\ref{eq:VectorDecompostion_1}) relies on performing vSFTs, either from eq. (\ref{eq:EdmondsvSFT}), or from eq. (\ref{eq:vSFT}) followed by an Edmonds transformation of each of the modes. Vector fields $\bu^{(1)}(\br)$ and $\bu^{(2)}(\br)$ are \emph{degree-dependent} linear combinations of radial and tangential fields. Therefore,  obtaining decomposition (\ref{eq:VectorDecompostion_1}) requires performing SFTs to determine the coefficients $\hat{u}^{(i)}_{n,j}(r)$ in eq. (\ref{eq:EdmondsVectors}).

\subsection{Generalizing the convolution to vector fields on $\cS$}
We now introduce the following generalized convolution operation on vector field $\bu\in$ {\bf l}$^q(\cS)$, 
\be
G_\bx\overrightarrow{*}\bu\coloneqq G_\bx^{(1)}*\bu^{(1)} + G_\bx^{(2)}*\bu^{(2)} + G_\bx^{(3)}*\bu^{(3)},
\lb{eq:GeneralVectorConv}\ee
where $G\in L^{p}[-1,1]$, with $p,q\in(1,\infty)$ and $1/p+1/q=1$. $G^{(1)}$ and $G^{(2)}$ are spectrally shifted versions of $G$,
\begin{eqnarray}
G_\bx^{(1)}(\br) &=& G^{(1)}\left(\frac{\bx\bdot\br}{r^2}\right)\coloneqq {\mathcal {LT}^{-1}}\{\widehat{G}(n-1)\},\\
G_\bx^{(2)}(\br) &=&G^{(2)}\left(\frac{\bx\bdot\br}{r^2}\right)\coloneqq  {\mathcal {LT}^{-1}}\{\widehat{G}(n+1)\},\\
G_\bx^{(3)}(\br) &\coloneqq& G\left(\frac{\bx\bdot\br}{r^2}\right).
\end{eqnarray}
We assign an arbitrary value to $\widehat{G}(-1) = \const$; our results here are independent of any particular choice. 

\begin{Lem} For a zonal function, $G_\bx$ on $\cS$, such that $G \in L^1[-1,1]$, the vector spherical harmonic basis functions, $\bY_{n,j}$, $\bPsi_{n,j}$, and $\bPhi_{n,j}$, are eigenfunctions of the generalized vector convolution, $G_\bx\overrightarrow{*}$, defined in eq. (\ref{eq:GeneralVectorConv}), with eigenvalue $\widehat{G}(n)$.
\lb{lem:VectorConvEigen}\end{Lem}
\vspace{.5cm}
\noindent {\it Proof of Lemma \ref{lem:VectorConvEigen}:}\\ 
\begin{eqnarray}
G_\bx\overrightarrow{*}\left(
\begin{array}{c}
\bY_{n,j}\\
\bPsi_{n,j}\\
\bPhi_{n,j}\\
\end{array}
\right)
&=&
[S^{-1}]\left(
\begin{array}{c}
G_\bx^{(1)}*\bK_{n,j}^{(1)}\\
G_\bx^{(2)}*\bK_{n,j}^{(2)}\\
G_\bx^{(3)}*\bK_{n,j}^{(3)}\\
\end{array}
\right)\nonumber\\
\mbox{from eq. (\ref{eq:ConvEdmondsBasis})~~~~}&=&
[S^{-1}]\left(
\begin{array}{c}
\widehat{G}(n)~\bK_{n,j}^{(1)}\\
\widehat{G}(n)~\bK_{n,j}^{(2)}\\
\widehat{G}(n)~\bK_{n,j}^{(3)}\\
\end{array}
\right)
=
\widehat{G}(n)
\left(\begin{array}{c}
\bY_{n,j}\\
\bPsi_{n,j}\\
\bPhi_{n,j}\\
\end{array}
\right)\nonumber
\end{eqnarray}
\hfill $\Box$
\\
\\
REMARK: Lemma \ref{lem:VectorConvEigen} is true by our design of the vector convolution, $G_\bx\overrightarrow{*}$. In fact, definition (\ref{eq:GeneralVectorConv}) was motivated by the requirement that such a generalized convolution satisfy Lemma \ref{lem:VectorConvEigen}. This is key to ensuring that differential operators commute with the generalized filtering operation, as we shall show below. It is worth noting that the Edmonds basis functions, $\bK_{n,j}^{(i)}$ for $i=1,2,3$, are also eigenfunctions of $G_\bx\overrightarrow{*}$, with eigenvalue $\widehat{G}(n)$, as is obvious from the proof above. However, our primary interest is in the canonical basis functions $\bY_{n,j}$, $\bPsi_{n,j}$, and $\bPhi_{n,j}$, which are more commonly used and have an intuitive geometric interpretation on the sphere.

\section{Convolution of Tensor Fields\lb{sec:Tensors}}
Here, we restrict ourselves to rank-2 tensors on $\cS$, and denote the dyadic (tensor) product of two vectors in $\cR^3$ by $\bx\otimes\by=\bx\by^{T}$. We also define \cite{FreedenSchreiner09} the operators $\grad^*\otimes$ and $\bL^*\otimes$, which act on a vector $\bu\in\cS$, as
\begin{eqnarray}
\grad^*\otimes\bu = \sum_{i=1}^3 \left(\grad^*u_i\right) \otimes \hat{\Be}_i\\
\bL^*\otimes\bu = \sum_{i=1}^3 \left(\bL^*u_i\right) \otimes \hat{\Be}_i
\end{eqnarray}
where $u_i = \bu\bdot\hat{\Be}_i$ is the $i$-th Cartesian vector component. 
An orthonormal basis for rank-2 tensors 
 \cite{FreedenSchreiner09} is:
\begin{eqnarray}
\bY^{(1,1)}_{n,j} &=& \hat{\Be}_r \otimes \bY_{n,j}\lb{eq:TensorBasis1}\\
\bY^{(1,2)}_{n,j} &=& \hat{\Be}_r \otimes \bPsi_{n,j}\lb{eq:TensorBasis2}\\
\bY^{(1,3)}_{n,j} &=& \hat{\Be}_r \otimes \bPhi_{n,j}\lb{eq:TensorBasis3}\\
\bY^{(2,1)}_{n,j} &=&  \bPsi_{n,j}\otimes\hat{\Be}_r \lb{eq:TensorBasis4}\\
\bY^{(2,2)}_{n,j} &=&  \frac{1}{\sqrt{2}}\left(\grad^*\otimes\bY_{n,j} -\grad^*Y_{n,j}\otimes\hat\Be_r\right) \lb{eq:TensorBasis5}\\
\bY^{(2,3)}_{n,j} &=&  \frac{1}{\sqrt{2\left(n\left(n+1\right)-2\right)}}\left(\grad^*\otimes\bPsi_{n,j} - \bL^*\otimes\bPhi_{n,j} + 2\bPsi_{n,j}\otimes\hat\Be_r\right)\lb{eq:TensorBasis6}\\
\bY^{(3,1)}_{n,j} &=&  \bPhi_{n,j}\otimes\hat{\Be}_r \lb{eq:TensorBasis7}\\
\bY^{(3,2)}_{n,j} &=& \frac{1}{\sqrt{2\left(n\left(n+1\right)-2\right)}}\left(\grad^*\otimes\bPhi_{n,j} + \bL^*\otimes\bPsi_{n,j} + 2\bPhi_{n,j}\otimes\hat\Be_r\right)\lb{eq:TensorBasis8}\\
\bY^{(3,3)}_{n,j} &=&  \frac{1}{\sqrt{2}}\left(\bL^*\otimes\bY_{n,j} -\bL^*Y_{n,j}\otimes\hat\Be_r\right)  \lb{eq:TensorBasis9}
\end{eqnarray}
The basis admits a geometrical interpretation as described in \cite{FreedenSchreiner09}: $\bY^{(1,1)}_{n,j}$ is normal to the sphere, 
$\bY^{(2,2)}_{n,j}$, $\bY^{(2,3)}_{n,j}$, $\bY^{(3,2)}_{n,j}$, and $\bY^{(3,3)}_{n,j}$ are tangential, $\bY^{(1,2)}_{n,j}$ and $\bY^{(1,3)}_{n,j}$ are normal-tangential, and $\bY^{(2,1)}_{n,j}$ and $\bY^{(3,1)}_{n,j}$ are tangential-normal.

Similar to the problem we had with the canonical vector basis functions (\ref{eq:VectorBasis1})-(\ref{eq:VectorBasis3}), tensorial basis functions $\bY^{(i,k)}_{n,j}$ for $i,k=1,2,3$ are not eigenfunctions of the traditional convolution $G_\bx*(\cdot)$. To this end, we will use an alternate orthonormal tensor basis \cite{Freedenetal98,FreedenSchreiner09},
\begin{eqnarray}
\bK^{(1,1)}_{n,j} &=& (\mbox{Norm. Fac.}) \left[\left((2n+3)\br - r^2\grad\right)\right.\lb{eq:AltTensorBasis1}\\
&& \hspace{2.5cm}\otimes\left.\left((2n+1)\br - r^2\grad\right) (r^n Y_{n,j})\right]_{r=1},\nonumber\\
\bK^{(1,2)}_{n,j} &=& (\mbox{Norm. Fac.}) \left[\left((2n-1)\br - r^2\grad\right)\otimes\left(\grad\right) (r^n Y_{n,j})\right]_{r=1},\lb{eq:AltTensorBasis2}\\
\bK^{(1,3)}_{n,j} &=& (\mbox{Norm. Fac.}) \left[\left((2n+1)\br - r^2\grad\right)\otimes\left(\br\btimes\grad\right) (r^n Y_{n,j})\right]_{r=1},\lb{eq:AltTensorBasis3}\\
\bK^{(2,1)}_{n,j} &=& (\mbox{Norm. Fac.}) \left[\grad \otimes\left((2n+1)\br - r^2\grad\right)(r^n Y_{n,j})\right]_{r=1},\lb{eq:AltTensorBasis4}\\
\bK^{(2,2)}_{n,j} &=& (\mbox{Norm. Fac.}) \left[\left(\grad\otimes\grad\right) (r^n Y_{n,j})\right]_{r=1},\lb{eq:AltTensorBasis5}\\
\bK^{(2,3)}_{n,j} &=& (\mbox{Norm. Fac.}) \left[\grad \otimes\left(\br\btimes\grad\right)(r^n Y_{n,j})\right]_{r=1},\lb{eq:AltTensorBasis6}\\
\bK^{(3,1)}_{n,j} &=& (\mbox{Norm. Fac.}) \left[\left(\br\btimes\grad\right) \otimes\left((2n+1)\br - r^2\grad\right)(r^n Y_{n,j})\right]_{r=1},\lb{eq:AltTensorBasis7}\\
\bK^{(3,2)}_{n,j} &=& (\mbox{Norm. Fac.}) \left[\left(\br\btimes\grad\right) \otimes\left(\grad\right)(r^n Y_{n,j})\right]_{r=1},\lb{eq:AltTensorBasis8}\\
\bK^{(3,3)}_{n,j} &=& (\mbox{Norm. Fac.}) \left[\left(\br\btimes\grad\right) \otimes\left(\br\btimes\grad\right)(r^n Y_{n,j})\right]_{r=1},\lb{eq:AltTensorBasis9}
\end{eqnarray}
where $(\mbox{Norm. Fac.})$ is a normalization factor (for details, see Ch. 6 in \cite{FreedenSchreiner09}). The advantage of this basis is that the Cartesian tensor components of each of $\bK^{(i,k)}_{n,j}$ for $i,k=1,2,3$ is a scalar spherical harmonic of degree $n\pm2$, $n\pm1$, and $n$ \cite{Freedenetal98}. The same is not true for basis (\ref{eq:TensorBasis1})-(\ref{eq:TensorBasis9}). This property makes $\bK^{(i,k)}_{n,j}$  eigenfunctions of the traditional convolution operation (\ref{eq:ConvolutionScalar}) (see \cite{FreedenSchreiner09} for details):
\begin{eqnarray}
G_\bx* \bK^{(i,k)}_{n,j} &=& \sum_{l,m=1}^3 G_\bx*\left(\bK^{(i,k)}_{n,j}\right)_{l,m}\hat{\Be}_{l}\otimes\hat{\Be}_{m}\nonumber\\
&=&\widehat{G}\left(\mbox{deg}\left(i,k \right)\right) \,\bK^{(i,k)}_{n,j}		\hspace{2cm} i,k=1,2,3
\lb{eq:ConvTensorBasis}\end{eqnarray}
where $\hat{\Be}_{m}$ for $m=1,2,3$ are the Cartesian unit vectors and $\left(\bK^{(i,k)}_{n,j}\right)_{l,m}$ is the $(l,m)$-th Cartesian tensor component. Moreover,
\begin{eqnarray}
\mbox{deg}\left(i,k \right) 
&=&\left\{
\begin{array}{ccl}
n-2 & ~~  &(i,k)=(2,2)\\
n-1  & ~~ &(i,k)=(2,3),(3,2)\\
n     & ~~ &(i,k)=(1,2),(2,1),(3,3)\\
n+1 & ~~ &(i,k)=(1,3),(3,1)\\
n+2 & ~~ &(i,k)=(1,1)\\
\end{array}\right.
\end{eqnarray}

The alternate tensor basis $\bK^{(i,k)}_{n,j}$ in eq. (\ref{eq:AltTensorBasis1})-(\ref{eq:AltTensorBasis9}) is related to basis $\bY^{(i,k)}_{n,j}$ in eq. (\ref{eq:TensorBasis1})-(\ref{eq:TensorBasis9}) through a linear transformation \cite{FreedenSchreiner09}:
\be
\left(
\begin{array}{c}
{\bK}^{(1,3)}_{n,j}\\
{\bK}^{(2,3)}_{n,j}\\
{\bK}^{(3,1)}_{n,j}\\
{\bK}^{(3,2)}_{n,j}\\
\end{array}
\right)
=
[B]\left(
\begin{array}{c}
 {\bY}^{(1,3)}_{n,j}\\
 {\bY}^{(3,1)}_{n,j}\\
 {\bY}^{(3,2)}_{n,j}\\
 {\bY}^{(3,3)}_{n,j}\\
\end{array}
\right) 
\hspace{.5cm}
\mbox{and}
\hspace{.5cm}
\left(
\begin{array}{c}
{\bK}^{(1,1)}_{n,j}\\
{\bK}^{(1,2)}_{n,j}\\
{\bK}^{(2,1)}_{n,j}\\
{\bK}^{(2,2)}_{n,j}\\
{\bK}^{(3,3)}_{n,j}\\
\end{array}
\right)
=
[A]\left(
\begin{array}{c}
 {\bY}^{(1,1)}_{n,j}\\
 {\bY}^{(1,2)}_{n,j}\\
 {\bY}^{(2,1)}_{n,j}\\
 {\bY}^{(2,2)}_{n,j}\\
 {\bY}^{(2,3)}_{n,j}\\
\end{array}
\right),
\lb{eq:EdmondsTranformTensors}\ee
where
\be [B]=
[E^{-1}_b]\left(
\begin{array}{cccrcr}
 n+1 & 1 &~& -\frac{1}{2} &~& -\frac{1}{2} n (n+1) \\
 n & -1 &~& \frac{1}{2} &~& \frac{1}{2} n (n+1) \\
 0 & n+2 &~& -\frac{1}{2} &~& \frac{1}{2} (n+1) (n+2) \\
 0 & n-1 &~& \frac{1}{2} &~& -\frac{1}{2} (n-1) n \\
\end{array}
\right)[C_b],
\ee
\begin{eqnarray} 
&&[A]=[E^{-1}_a]\nonumber\\
&&\left(
\begin{array}{c c c c c c c r}
 (n+1) (n+2) &~& -(n+2) &~& -(n+2) &~& \frac{1}{2} (-n-2) (n+1) & \frac{1}{2} \\
 n^2 &~& n &~& 1-n &~& \frac{1}{2} (n-1) n & -\frac{1}{2} \\
 (n+1)^2 &~& -(n+1) &~& n+2 &~& \frac{1}{2} (n+1) (n+2) & -\frac{1}{2} \\
 (n-1) n &~& n-1 &~& n-1 &~& -\frac{1}{2} (n-1) n & \frac{1}{2} \\
 0 &~& 0 &~& 1 &~& -\frac{1}{2} n (n+1) & -\frac{1}{2} \\
\end{array}
\right)[C_a],\hspace{-1cm}
\end{eqnarray} 
\be [C_b]=
\left(
\begin{array}{cccc}
 \sqrt{n (n+1)} & 0 & 0 & 0 \\
 0 & \sqrt{n (n+1)} & 0 & 0 \\
 0 & 0 &  \sqrt{2n (n+1) (n (n+1)-2)} & 0 \\
 0 & 0 & 0 & \sqrt{2} \\
\end{array}
\right),
\ee
\be [E_b]=
(2n+1)^{1\over2}\left(
\begin{array}{cccc}
 \sqrt{n (n+1)^2} & 0 & 0 & 0 \\
 0 & \sqrt{n^2 (n+1)} & 0 & 0 \\
 0 & 0 & \sqrt{n (n+1)^2} & 0 \\
 0 & 0 & 0 & \sqrt{(n-1) n^2} \\
\end{array}
\right),
\ee
\be [C_a]=
\left(
\begin{array}{ccccc}
 1 & 0 & 0 & 0 & 0 \\
 0 & \sqrt{n (n+1)} & 0 & 0 & 0 \\
 0 & 0 & \sqrt{n (n+1)} & 0 & 0 \\
 0 & 0 & 0 & \sqrt{2} & 0 \\
 0 & 0 & 0 & 0 &\sqrt{2n (n+1) (n (n+1)-2)} \\
\end{array}
\right),
\ee
\begin{eqnarray}
[E_a]&=&
(2n-1)^{1\over2}(2n-3)^{1\over2}\mbox{diag}
\left\{\sqrt{(n+1) (n+2)},\sqrt{\frac{3n^4}{ (2 n-3) (2 n-1)}},\right.\nonumber\\
&& \left.\sqrt{(n+1)^2},\sqrt{\frac{(n-1) n (2 n+1)}{ (2 n-3)}},\sqrt{\frac{n^2 (n+1) (n+2)}{ (2 n-3) (2 n-1)}}\right\}.
\end{eqnarray}

A tensor $\bT\in$ $\bL^{2}(\cS)$ can be represented as a Fourier series:
\begin{eqnarray}
\bT(\br) &=& \sum_{n=0}^\infty \sum_{j=-n}^n \left(\sum_{i,k=1}^3 \hat{T}^{Y^{(i,k)}}_{n,j}(r) ~~ \bY^{(i,k)}_{n,j}  \right)\lb{eq:TensorSFT_1}\\
&=& \sum_{n=0}^\infty \sum_{j=-n}^n   \left(\sum_{i,k=1}^3 \hat{T}^{(i,k)}_{n,j}(r) ~~\bK^{(i,k)}_{n,j} ~ ~\right)\lb{eq:TensorSFT_2}
\end{eqnarray}
where the Fourier coefficients are obtained from the following tensor Spherical Fourier Transform (tSFT):
\begin{eqnarray}
\hat{T}^{Y^{(i,k)}}_{n,j}(r) &=&\langle\bT\bdot\bY^{(i,k)}_{n,j}\rangle, \hspace{.5cm}i,k=1,2,3\lb{eq:tSFT}\\
\hat{T}^{(i,k)}_{n,j}(r) &=&\langle\bT\bdot \bK^{(i,k)}_{n,j} \rangle, \hspace{.5cm}i,k=1,2,3\lb{eq:EdmondstSFT}
\end{eqnarray}
where the $(\bdot)$ in eqs. (\ref{eq:tSFT}),(\ref{eq:EdmondstSFT}) is a tensor scalar product.
In light of this, a tensor field may be decomposed into nine parts:
\be \bT(\br) = \sum_{i,k=1}^3\bT^{(i,k)}(\br)
\lb{eq:TensorDecompostion_1}\ee
where
\be
\bT^{(i,k)}(\br) = \sum_{n=0}^\infty \sum_{j=-n}^n \hat{T}^{(i,k)}_{n,j}(r)  ~~\bK^{(i,k)}_{n,j}, \hspace{1cm} i,k=1,2,3
\lb{eq:EdmondsTensors}\ee
Note that the decomposition in eq.(\ref{eq:TensorDecompostion_1}) relies on performing tensor SFTs, either through eq. (\ref{eq:EdmondstSFT}), or through eq. (\ref{eq:tSFT}) followed by linear transformation (\ref{eq:EdmondsTranformTensors}) of each of the modes to obtain the term on the left-hand-side of eq. (\ref{eq:EdmondsTensors}).

\subsection{Generalizing the convolution to rank-2 tensor fields on $\cS$}
Similar to the vector convolution, we also introduce the tensor convolution on tensor field $\bT\in$ $\bL^{q}(\cS)$,
\be
G_\bx\overleftrightarrow{*}\bT\coloneqq \sum_{k=1}^3\sum_{i=1}^3 G_\bx^{(i,k)}*\bT^{(i,k)} \,,
\lb{eq:GeneralTensorConv}\ee
where $\bx\in\cS$ and $G\in L^{p}[-1,1]$ such that $p,q\in(1,\infty)$ and $1/p+1/q=1$. $G_\bx^{(i,k)}$ are spectrally shifted versions of $G$,
\begin{eqnarray}
G_\bx^{(1,1)}(\br) &\coloneqq& {\mathcal {LT}^{-1}}\{\widehat{G}(n-2)\},\\
G_\bx^{(2,2)}(\br) &\coloneqq&  {\mathcal {LT}^{-1}}\{\widehat{G}(n+2)\},\\
G_\bx^{(2,3)}(\br)=G_\bx^{(3,2)}(\br) &\coloneqq&  {\mathcal {LT}^{-1}}\{\widehat{G}(n+1)\},\\
G_\bx^{(1,3)}(\br)=G_\bx^{(3,1)}(\br) &\coloneqq&  {\mathcal {LT}^{-1}}\{\widehat{G}(n-1)\},\\
G_\bx^{(1,2)}(\br)=G_\bx^{(2,1)}(\br)=G_\bx^{(3,3)}(\br) &\coloneqq& G.
\end{eqnarray}
We assign arbitrary values to $\widehat{G}(-1)$ and $\widehat{G}(-2)$; our results here are independent of any particular choice.

\begin{Lem} For a zonal function, $G_\bx$ on $\cS$, such that $G \in L^1[-1,1]$, the tensor spherical harmonic basis functions, $\bY^{(i,k)}_{n,j}$ for $i,k=1,2,3$, are eigenfunctions of the generalized tensor convolution, $G_\bx\overleftrightarrow{*}$, defined in eq. (\ref{eq:GeneralTensorConv}), with eigenvalue $\widehat{G}(n)$.
\lb{lem:TensorConvEigen}\end{Lem}
\vspace{.5cm}
\noindent {\it Proof of Lemma \ref{lem:TensorConvEigen}:}
\begin{eqnarray}
G_\bx\overleftrightarrow{*}\left(
\begin{array}{c}
\bY^{(1,3)}_{n,j}\\
\bY^{(3,1)}_{n,j}\\
\bY^{(3,2)}_{n,j}\\
\bY^{(3,3)}_{n,j}\\
\end{array}
\right)
&=&
[B^{-1}]\left(
\begin{array}{c}
G_\bx^{(1,3)}*\bK_{n,j}^{(1,3)}\\
G_\bx^{(2,3)}*\bK_{n,j}^{(2,3)}\\
G_\bx^{(3,1)}*\bK_{n,j}^{(3,1)}\\
G_\bx^{(3,2)}*\bK_{n,j}^{(3,2)}\\
\end{array}
\right)\nonumber\\
\mbox{from eq. (\ref{eq:ConvTensorBasis})~~~~}&=&
[B^{-1}]\left(
\begin{array}{c}
\widehat{G}(n)~\bK_{n,j}^{(1,3)}\\
\widehat{G}(n)~\bK_{n,j}^{(2,3)}\\
\widehat{G}(n)~\bK_{n,j}^{(3,1)}\\
\widehat{G}(n)~\bK_{n,j}^{(3,2)}\\
\end{array}
\right)
=
\widehat{G}(n)
\left(\begin{array}{c}
\bY^{(1,3)}_{n,j}\\
\bY^{(3,1)}_{n,j}\\
\bY^{(3,2)}_{n,j}\\
\bY^{(3,3)}_{n,j}\\
\end{array}
\right)\nonumber
\end{eqnarray}

Similarly, we have
\begin{eqnarray}
&&G_\bx\overleftrightarrow{*}\left(\bY^{(1,1)}_{n,j}~\bY^{(1,2)}_{n,j}~\bY^{(2,1)}_{n,j}~\bY^{(2,2)}_{n,j}~\bY^{(2,3)}_{n,j}\right)^T\nonumber\\
= &&\widehat{G}(n)~\left(\bY^{(1,1)}_{n,j}~\bY^{(1,2)}_{n,j}~\bY^{(2,1)}_{n,j}~\bY^{(2,2)}_{n,j}~\bY^{(2,3)}_{n,j}\right)^T.
\nonumber\end{eqnarray}

\hfill $\Box$
\\
\\
REMARK: Similar to Lemma \ref{lem:VectorConvEigen}, Lemma \ref{lem:TensorConvEigen} is also true by design of the tensor convolution, $G_\bx\overleftrightarrow{*}$. It is key to ensuring that the generalized filtering operation commutes with spatial derivative operators in PDEs. Tensors $\bK_{n,j}^{(i,k)}$ for $i,k=1,2,3$, are also eigenfunctions of $G_\bx\overleftrightarrow{*}$, with eigenvalue $\widehat{G}(n)$. However, our primary interest is in the canonical basis functions $\bY^{(i,k)}_{n,j}$ for $i,k=1,2,3$, which have an intuitive geometric interpretation on the sphere.

\section{Spherical Filtering\lb{sec:SphericalFiltering}}
As mentioned in the Introduction, we need to generalize the filtering (or coarse-graining) operation to allow for a scale-decomposition on $\cS$. As a matter of terminology, we shall use the word {\it filter} in the sense of a {\it low-pass filter} which filters out the small-scale (or high-frequency) spatial variations of a signal. For $G\in L^p[-1,1]$, we define the following filtering operation on scalar field, $f\in L^q(\cS)$, vector field, $\bu\in$ {\bf l}$^q(\cS)$, and tensor field, $\bT\in$ $\bL^q(\cS)$, where $p,q\in(1,\infty)$ and $1/p+1/q=1$,
\begin{eqnarray}
\OL{f}(\bx) &\coloneqq& G_\bx*f \lb{eq:FilteringScalar}\\
\OL{\bu}(\bx) &\coloneqq& G_\bx\overrightarrow{*}\bu\lb{eq:FilteringVector}\\
\OL{\bT}(\bx) &\coloneqq& G_\bx\overleftrightarrow{*}\bT\lb{eq:FilteringTensor}.
\end{eqnarray}
Note that each of the convolutions in eqs. (\ref{eq:FilteringScalar})-(\ref{eq:FilteringTensor}) is consistent with the physical notion of filtering out variations at spatial scales smaller than the width of kernel $G$. 
The filtering operation, by its definition in eqs. (\ref{eq:FilteringScalar})-(\ref{eq:FilteringTensor}), is sensitive to whether the quantity being filtered is a scalar, vector, or tensor. For example, $\OL{\grad f} = G_\bx\overrightarrow{*}(\grad f)$, whereas $\grad \OL{f} = \grad (G_\bx*f)$.

\begin{Property} The filtering operation defined in eqs. (\ref{eq:FilteringScalar})-(\ref{eq:FilteringTensor}) is linear.
\lb{lem:FilterLinearity}\end{Property}
The proof is straightforward and follows from linearity of the traditional convolution in eq. (\ref{eq:ConvolutionScalar}) and its generalization to vectors and tensors.
\\
\\
REMARK: An important consequence of Property \ref{lem:FilterLinearity} is that for a normalized zonal kernel, with $G\in L^1[-1,1]$,
and spatially uniform scalar field $f_0$, vector field $\bu_0$, and tensor field $\bT_0$ on $\cS$, we have 
$$  \OL{f_0}  =  f_0,	\hspace{.5cm}
 \OL{\bu_0}  =  \bu_0,	\hspace{.5cm}
 \OL{\bT_0}  =  \bT_0.
$$
Here, a spatially uniform vector or tensor field means one whose Cartesian components are independent of position on $\cS$. This shows that our $\OL{\left(\cdot\right)}$ operation satisfies the physical expectation that uniform fields, by definition, do not vary at scales smaller than that of the domain and, as a result, cannot be altered by a low-pass filter.

\begin{Property} The filtering operation defined in eqs. (\ref{eq:FilteringScalar})-(\ref{eq:FilteringTensor}) is ~~~`mean-preserving': For a normalized zonal kernel $G_\bx$ on $\cS$ such that $G\in L^p[-1,1]$, and scalar field, $f\in L^q(\cS)$, vector field, $\bu\in$ {\bf l}$^q(\cS)$, and tensor field, $\bT\in$ $\bL^q(\cS)$, where $p,q\in(1,\infty)$ and $1/p+1/q=1$, we have 
$$ \langle \OL{f} \rangle = \langle f \rangle,	\hspace{.5cm}
\langle \OL\bu \rangle = \langle \bu \rangle	,	\hspace{.5cm}
\langle \OL\bT \rangle = \langle \bT \rangle,
$$
where $\langle\dots\rangle$, defined in eq. (\ref{eq:def_average}), is the space-average on $\cS$.
\lb{lem:FilterMean}\end{Property}
\vspace{.5cm}
\noindent {\it Proof of Property \ref{lem:FilterMean}:}\\
The statement for scalars follows directly from the definition of convolution:
\begin{eqnarray}
\langle \OL{f} \rangle&=&\frac{1}{4\pi r^2}\int_\cS dS(\bx) \int_\cS dS(\br)~G(\frac{\bx\bdot\br}{r^2})f(\br) \nonumber\\
&=&\frac{1}{4\pi r^2}\int_\cS dS(\br)~f(\br)\left(\int_\cS dS(\bx)~G(\frac{\bx\bdot\br}{r^2})\right) 
=\langle {f} \rangle \nonumber
\end{eqnarray}
The statement for vectors builds upon that for scalars:
\begin{eqnarray}
\langle \OL{\bu} \rangle&=&\sum_{i=1}^3\langle G_\bx^{(i)}*\bu^{(i)} \rangle \hspace{1cm}\mbox{using definition (\ref{eq:GeneralVectorConv})}\nonumber\\
&=&\sum_{i=1}^3 \sum_{m=1}^3 \langle G_\bx^{(i)}*u_m^{(i)}\rangle\hat{\Be}_m
=\sum_{i=1}^3 \sum_{m=1}^3 \langle u_m^{(i)}\rangle\hat{\Be}_m =\langle {\bu} \rangle~~.\nonumber
\end{eqnarray}
The statement for tensors follows from a similar proof.
\hfill $\Box$
\\
\\
REMARK: Property \ref{lem:FilterMean} ensures that our filtering operation does not alter the mean values of fields. It simply redistributes (or smears) the field in space (over $\cS$) as should be expected from any filter.

\section{Differential operators and filtering commute\lb{sec:FilteringDerivativesCommute}}
In this section, we present the main results of this paper. We shall prove that spherical filtering, as defined in eqs. (\ref{eq:FilteringScalar})-(\ref{eq:FilteringTensor}), commutes with differential operators  on $\cS$.
\begin{Prop} For a zonal kernel $G_\bx$ on $\cS$, such that $G\in L^p[-1,1]$, $f\in L^q(\cS)$, vector field, $\bu\in$ {\bf l}$^q(\cS)$, and tensor field, $\bT\in$ $\bL^q(\cS)$, with $p,q\in(1,\infty)$ and $1/p+1/q=1$, and using the spherical filtering operation, $\OL{\left(\dots\right)}$, defined in eqs. (\ref{eq:FilteringScalar})-(\ref{eq:FilteringTensor}), we have,
\begin{eqnarray}
\mbox{assuming}~ f\in C^{(1)}(\cS), \hspace{1.7cm}\OL{\grad^* f} &=& \grad^* \OL{f}~,\lb{eq:FilterGradHozScalarCommute}\\
\mbox{assuming}~ f\in C^{(1)}(\cS), \hspace{1.8cm}\OL{\bL^* f} &=& \bL^* \OL{f}~, \lb{eq:FilterCurlHozScalarCommute}\\
\OL{f \,\hat{\Be}_r} &=& \OL{f} \,\hat{\Be}_r~,\lb{eq:FilterScalarRadialUnitVectorCommute}\\
\OL{\bu \bdot\hat{\Be}_r} &=& \OL{\bu} \bdot \hat{\Be}_r~,\lb{eq:FilterVectorDotRadialUnitVectorCommute}\\
\OL{\bu \btimes\hat{\Be}_r} &=& \OL{\bu} \btimes \hat{\Be}_r~,\lb{eq:FilterVectorCrossRadialUnitVectorCommute}\\
\mbox{assuming}~ \bu\in{\bf c}^{(1)}(\cS), \hspace{1.5cm}\OL{\grad^* \bdot\bu} &=& \grad^* \bdot\OL{\bu}~,\lb{eq:FilterDivHozVectorCommute}\\
\mbox{assuming}~ \bu\in{\bf c}^{(1)}(\cS), \hspace{1.6cm}\OL{\bL^* \bdot\bu} &=& \bL^* \bdot\OL{\bu},\lb{eq:FilterCurlDotHozVectorCommute}\\
\mbox{assuming}~ \bu\in{\bf c}^{(1)}(\cS), \hspace{1.25cm}\OL{\grad^* \btimes\bu} &=& \grad^* \btimes\OL{\bu}~,\lb{eq:FilterRadialCurlVectorCommute}\\
\mbox{assuming}~ f\in C^{(2)}(\cS), \hspace{1.6cm} \OL{\Delta^* f} &=& \Delta^*\OL{f}~,\lb{eq:FilterLaplaceHozScalarCommute}\\
\OL{\hat{\Be}_r \otimes\bu} &=&  \hat{\Be}_r \otimes\OL{\bu}~,\lb{eq:FilterTensorRadialUnitVectorVectorCommute}\\
\OL{ \bu\otimes\hat{\Be}_r} &=&  \OL{\bu}\otimes\hat{\Be}_r~, \lb{eq:FilterTensorVectorRadialUnitVectorCommute}\\
\mbox{assuming}~ \bu\in{\bf c}^{(1)}(\cS), \hspace{1cm}\OL{\grad^* \otimes\bu} &=&  \grad^*\otimes \OL{\bu}~,\lb{eq:FilterTensorGradHozVectorCommute}\\
\mbox{assuming}~ \bu\in{\bf c}^{(1)}(\cS), \hspace{1.1cm}\OL{\bL^* \otimes\bu} &=& \bL^*\otimes \OL{\bu}~,\lb{eq:FilterTensorCurlHozVectorCommute}\\
\mbox{assuming}~ \bT\in{\bf C}^{(1)}(\cS), \hspace{1.2cm}\OL{\grad^*\bdot\bT} &=& \grad^*\bdot\OL{\bT}~,\lb{eq:FilterDivHozTensorCommute}\\
\mbox{assuming}~ \bT\in{\bf C}^{(1)}(\cS), \hspace{1.25cm}\OL{\bL^*\bdot\bT} &=& \bL^*\bdot\OL{\bT}~.\lb{eq:FilterCurlDotHozTensorCommute}
\end{eqnarray}
\lb{prop:FilterHozDerivCommute}\end{Prop}
\vspace{.5cm}
\noindent {\it Proof of Proposition \ref{prop:FilterHozDerivCommute}:}\\
We have the following Fourier decompositions:
\begin{eqnarray}
\grad^* f &=& \sum_{n,j} \hat{f}_{n,j} \sqrt{n(n+1)}~\bPsi_{n,j},\nonumber\\
\bL^* f &=& \sum_{n,j} \hat{f}_{n,j} \sqrt{n(n+1)}~\bPhi_{n,j}, \nonumber\\
\bu&=& \sum_{n,j} \hat{u}^{Y}_{n,j} \bY_{n,j} + \hat{u}^{\psi}_{n,j} \bPsi_{n,j} + \hat{u}^{\phi}_{n,j} \bPhi_{n,j}\nonumber\\
\bT &=& \sum_{n=0}^\infty \sum_{j=-n}^n \left(\sum_{i,k=1}^3 \hat{T}^{Y^{(i,k)}}_{n,j}(r) ~~ \bY^{(i,k)}_{n,j}  \right)\nonumber\end{eqnarray}

Relation (\ref{eq:FilterGradHozScalarCommute}) follows from:
\begin{eqnarray}
\OL{\grad^* f} 
&=& G_\bx\overrightarrow{*}\sum_{n,j} \hat{f}_{n,j} \sqrt{n(n+1)}~\bPsi_{n,j}\nonumber\\
&=& \sum_{n,j} \hat{f}_{n,j} \sqrt{n(n+1)}~\widehat{G}(n)~\bPsi_{n,j}\nonumber\\
&=& \grad^*\sum_{n,j} \hat{f}_{n,j} \widehat{G}(n)~Y_{n,j}\nonumber\\
&=& \grad^*(G_\bx*f)\nonumber
\end{eqnarray}
Relation (\ref{eq:FilterCurlHozScalarCommute}) follows from similar considerations.

Relation (\ref{eq:FilterScalarRadialUnitVectorCommute}) follows from the fact that
$\bw\equiv f \,\hat{\Be}_r$ is purely radial. Therefore, $\bw = \sum_{n,j} \hat{f}_{n,j}(r)\,\bY_{n,j}$.
We then have
\begin{eqnarray}
\OL{\bw} 
&=& \sum_{n,j} \hat{f}_{n,j}(r)\,G_\bx\overrightarrow{*}\,\bY_{n,j}\nonumber\\
\text{using Lemma \ref{lem:VectorConvEigen}}\hspace{.5cm}
&=& \sum_{n,j} \hat{f}_{n,j}(r)\,\widehat{G}(n)\,\bY_{n,j}\nonumber\\
&=& \hat{\Be}_r \sum_{n,j} \hat{f}_{n,j}(r)\,(G_\bx*\,Y_{n,j}) \nonumber\\
&=& \hat{\Be}_r\, G_\bx*\sum_{n,j} \hat{f}_{n,j}(r)\,Y_{n,j} \nonumber\\
&=& \hat{\Be}_r \,(G_\bx*f) =  \OL{f}\,\hat{\Be}_r\nonumber
\end{eqnarray}

Relation (\ref{eq:FilterVectorDotRadialUnitVectorCommute}) follows from the following considerations.
For a purely tangential vector field, $\bu_t = \sum_{n=0}^{\infty}\sum_{j=-n}^{n} \hat{u}^{(\psi)}_{n,j}(r)\bPsi_{n,j}+  \hat{u}^{(\phi)}_{n,j}(r)\bPhi_{n,j}\nonumber$, we have
\begin{eqnarray}
\OL{\bu_t}=G_\bx\overrightarrow{*}\bu_t 
&=& \sum_{n=0}^{\infty}\sum_{j=-n}^{n} \hat{u}^{(\psi)}_{n,j}(r)\,(G_\bx\overrightarrow{*} \bPsi_{n,j})+  \hat{u}^{(\phi)}_{n,j}(r)\,(G_\bx\overrightarrow{*} \bPhi_{n,j})\hspace{3cm}\nonumber\\
\text{using Lemma \ref{lem:VectorConvEigen}}\hspace{.5cm}
&=& \sum_{n=0}^{\infty}\sum_{j=-n}^{n} \hat{u}^{(\psi)}_{n,j}(r)\,\widehat{G}(n)\, \bPsi_{n,j}+  \hat{u}^{(\phi)}_{n,j}(r)\,\widehat{G}(n)\,\bPhi_{n,j}\nonumber
\end{eqnarray}
which is a tangential vector field. Therefore, for $\bu= u_r \hat{\Be}_r + \bu_t$
\begin{eqnarray}
\OL{\bu} \bdot \hat{\Be}_r&=& \left(\OL{u_r \,\hat{\Be}_r} + \OL{\bu_t}\right)\bdot \hat{\Be}_r\nonumber\\
 &=& \left(\OL{u_r \,\hat{\Be}_r}\right)\bdot \hat{\Be}_r  + 0\nonumber\\
 &=& \OL{u_r}  + 0 = \OL{\bu\bdot\hat{\Be}_r}\nonumber,
\end{eqnarray}
where we used relation (\ref{eq:FilterScalarRadialUnitVectorCommute}) on the last line.

To prove relation (\ref{eq:FilterVectorCrossRadialUnitVectorCommute}) we first observe,
\begin{eqnarray}
\bY_{n,j}\btimes\hat{\Be}_r=\bzed, \hspace{.5cm} 
\bPsi_{n,j}\btimes\hat{\Be}_r=-\bPhi_{n,j}, \hspace{.5cm} 
\bPhi_{n,j}\btimes\hat{\Be}_r=\bPsi_{n,j}.
\end{eqnarray}
Therefore,
\begin{eqnarray}
\OL{\bu\btimes\hat{\Be}_r} 
&=& G_\bx\overrightarrow{*}\sum_{n,j} -\hat{u}^{\psi}_{n,j}\bPhi_{n,j}+\hat{u}^{\phi}_{n,j}\bPsi_{n,j}\nonumber\\
&=& \left(\sum_{n,j} \hat{u}^{Y}_{n,j}\,\widehat{G}(n)\,\bY_{n,j} + \hat{u}^{\psi}_{n,j}\,\widehat{G}(n)\,\bPsi_{n,j}+\hat{u}^{\phi}_{n,j}\,\widehat{G}(n)\,\bPhi_{n,j}\right)\btimes\hat{\Be}_r\nonumber\\
&=& \OL{\bu}\btimes\hat{\Be}_r \nonumber
\end{eqnarray}

Relation (\ref{eq:FilterDivHozVectorCommute}) follows from:
\begin{eqnarray}
\OL{\grad^* \bdot\bu}
&=& G_\bx*\sum_{n,j} \hat{u}^{\psi}_{n,j} \grad^* \bdot\bPsi_{n,j} \nonumber\\
&=& G_\bx*\sum_{n,j}  \frac{\hat{u}^{\psi}_{n,j}}{\sqrt{n(n+1)}}\Delta^* Y_{n,j} \nonumber\\
&=& -\sum_{n,j}  \frac{\hat{u}^{\psi}_{n,j}}{\sqrt{n(n+1)}} n(n+1)~G_\bx*Y_{n,j} \nonumber\\
&=& -\sum_{n,j}  \frac{\hat{u}^{\psi}_{n,j}}{\sqrt{n(n+1)}} n(n+1)~\widehat{G}(n)~Y_{n,j} \nonumber\\
&=& \grad^* \bdot\sum_{n,j}  \hat{u}^{\psi}_{n,j} ~\widehat{G}(n)~\bPsi_{n,j} \nonumber\\
&=& \grad^* \bdot\sum_{n,j}  \hat{u}^{Y}_{n,j} ~\widehat{G}(n)~\bY_{n,j}+\hat{u}^{\psi}_{n,j} ~\widehat{G}(n)~\bPsi_{n,j} + \hat{u}^{\phi}_{n,j} ~\widehat{G}(n)~\bPhi_{n,j} \nonumber\\
&=& \grad^* \bdot\left(G_\bx\overrightarrow{*}\sum_{n,j}  \hat{u}^{Y}_{n,j} \bY_{n,j}+\hat{u}^{\psi}_{n,j} \bPsi_{n,j}+ \hat{u}^{\phi}_{n,j} \bPhi_{n,j}\right) \nonumber\\
&=& \grad^* \bdot\OL{\bu} \nonumber
\end{eqnarray}
Relation (\ref{eq:FilterCurlDotHozVectorCommute}) follows from a similar analysis to that of relation (\ref{eq:FilterDivHozVectorCommute}).

Relation (\ref{eq:FilterRadialCurlVectorCommute}) follows from the fact that
$\grad^*\btimes\bu = \left(\bL^*\bdot\bu\right)\hat{\Be}_r$, along with
results (\ref{eq:FilterScalarRadialUnitVectorCommute}) and (\ref{eq:FilterCurlDotHozVectorCommute}).

Relation (\ref{eq:FilterLaplaceHozScalarCommute}) follows from 
$\Delta^* f =\grad^*\bdot\grad^* f = \bL^*\bdot\bL^* f $, along with results (\ref{eq:FilterDivHozVectorCommute}) and (\ref{eq:FilterGradHozScalarCommute}) or results (\ref{eq:FilterCurlDotHozVectorCommute}) and (\ref{eq:FilterCurlHozScalarCommute}).

The proof of relation (\ref{eq:FilterTensorRadialUnitVectorVectorCommute}) goes as follows,
\begin{eqnarray}
\OL{\hat{\Be}_r \otimes\bu} 
&=& G_\bx\overleftrightarrow{*}\hat{\Be}_r \otimes \sum_{n,j} \hat{u}^{Y}_{n,j}\bY_{n,j} + \hat{u}^{\psi}_{n,j}\bPsi_{n,j}+ \hat{u}^{\phi}_{n,j}\bPhi_{n,j}\nonumber\\
&=& \sum_{n,j} \hat{u}^{Y}_{n,j}~\widehat{G}(n)~\bY^{(1,1)}_{n,j} + \hat{u}^{\psi}_{n,j}~\widehat{G}(n)~\bY^{(1,2)}_{n,j}+ \hat{u}^{\phi}_{n,j}~\widehat{G}(n)~\bY^{(1,3)}_{n,j}\nonumber\\
&=& \hat{\Be}_r \otimes\sum_{n,j} \hat{u}^{Y}_{n,j}~\widehat{G}(n)~\bY_{n,j} + \hat{u}^{\psi}_{n,j}~\widehat{G}(n)~\bPsi_{n,j}+ \hat{u}^{\phi}_{n,j}~\widehat{G}(n)~\bPhi_{n,j}\nonumber\\
&=& \hat{\Be}_r \otimes\OL{\bu}\nonumber
\end{eqnarray}
Relation (\ref{eq:FilterTensorVectorRadialUnitVectorCommute}) follows a similar proof.

To prove relations (\ref{eq:FilterTensorGradHozVectorCommute}) and (\ref{eq:FilterTensorCurlHozVectorCommute}) we need the six identities
\begin{eqnarray}
\grad^*\otimes\grad^* Y_{n,j}&=& \frac{\sqrt{2(n(n+1)-2)}}{2} \bY^{(2,3)}_{n,j} -\bY^{(2,1)}_{n,j} - \frac{\sqrt{2}}{2}n(n+1)\bY^{(2,2)}_{n,j}\\
\bL^*\otimes\bL^* Y_{n,j}&=& -\frac{\sqrt{2(n(n+1)-2)}}{2} \bY^{(2,3)}_{n,j} +\bY^{(2,1)}_{n,j} - \frac{\sqrt{2}}{2}n(n+1)\bY^{(2,2)}_{n,j}\\
\grad^*\otimes\bL^* Y_{n,j}&=& \frac{\sqrt{2(n(n+1)-2)}}{2} \bY^{(3,2)}_{n,j} -\bY^{(3,1)}_{n,j} + \frac{\sqrt{2}}{2}n(n+1)\bY^{(3,3)}_{n,j}\\
\bL^*\otimes\grad^* Y_{n,j}&=& \frac{\sqrt{2(n(n+1)-2)}}{2} \bY^{(3,2)}_{n,j} -\bY^{(3,1)}_{n,j} - \frac{\sqrt{2}}{2}n(n+1)\bY^{(3,3)}_{n,j}\\
\grad^*\otimes Y_{n,j} \hat{\Be}_r&=& \sqrt{2}~\bY^{(2,2)}_{n,j} +\bY^{(2,1)}_{n,j}\\
\bL^*\otimes Y_{n,j} \hat{\Be}_r&=& \sqrt{2}~\bY^{(3,3)}_{n,j} +\bY^{(3,1)}_{n,j}.
\end{eqnarray}
The last two identities, along with accompanying derivations, can be found as eqs. (6.49)-(6.50) in \cite{FreedenSchreiner09}. The first four identities can be derived from eqs. (6.115)-(6.116) in \cite{FreedenSchreiner09} along with the above definitions (\ref{eq:TensorBasis1})-(\ref{eq:TensorBasis9}) of the tensor basis functions. From Lemma \ref{lem:TensorConvEigen}, we have
\begin{eqnarray}
&&\OL{\grad^*\otimes\bPsi_{n,j} }\\
&=& \frac{\widehat{G}(n)}{\sqrt{n(n+1)}}\left[\frac{\sqrt{2(n(n+1)-2)}}{2} \bY^{(2,3)}_{n,j} -\bY^{(2,1)}_{n,j} - \frac{\sqrt{2}}{2}n(n+1)\bY^{(2,2)}_{n,j}	\right]\nonumber\\
&=& \frac{\widehat{G}(n)}{\sqrt{n(n+1)}}\grad^*\otimes\grad^* Y_{n,j} = \grad^*\otimes \left[\widehat{G}(n) \bPsi_{n,j}\right] = \grad^*\otimes \OL{\bPsi_{n,j}} \nonumber
\end{eqnarray}
Similar considerations yield
\begin{eqnarray}
\OL{\bL^*\otimes\bPhi_{n,j}} &=& \bL^*\otimes \OL{\bPhi_{n,j}}~,\\
\OL{\grad^*\otimes\bPhi_{n,j}} &=& \grad^*\otimes \OL{\bPhi_{n,j}}~,\\
\OL{\bL^*\otimes\bPsi_{n,j}} &=& \bL^*\otimes \OL{\bPsi_{n,j}}~,\\
\OL{\grad^*\otimes\bY_{n,j}} &=& \grad^*\otimes\OL{\bY_{n,j}}~,\\
\OL{\bL^*\otimes\bY_{n,j}} &=& \bL^*\otimes\OL{\bY_{n,j}}~.
\end{eqnarray}
We finally get relations (\ref{eq:FilterTensorGradHozVectorCommute}) and (\ref{eq:FilterTensorCurlHozVectorCommute}) from
\begin{eqnarray}
\OL{\grad^*\otimes\bu} &=&  \sum_{n,j}\hat{\bu}_{n,j}^{Y} \OL{\grad^*\otimes \bY_{n,j}}  + \hat{\bu}_{n,j}^{\psi} \OL{\grad^*\otimes \bPsi_{n,j}} + \hat{\bu}_{n,j}^{\phi} \OL{\grad^*\otimes \bPhi_{n,j}}\nonumber\\
&=&  \sum_{n,j}\hat{\bu}_{n,j}^{Y} \grad^*\otimes \OL{\bY_{n,j}} +\hat{\bu}_{n,j}^{\psi} \grad^*\otimes \OL{\bPsi_{n,j}} + \hat{\bu}_{n,j}^{\phi} \grad^*\otimes \OL{\bPhi_{n,j}}
=  \grad^*\otimes\OL{\bu}\nonumber\\
\OL{\bL^*\otimes\bu} &=&  \sum_{n,j}\hat{\bu}_{n,j}^{Y} \OL{\bL^*\otimes \bY_{n,j}} +\hat{\bu}_{n,j}^{\psi} \OL{\bL^*\otimes \bPsi_{n,j}} + \hat{\bu}_{n,j}^{\phi} \OL{\bL^*\otimes \bPhi_{n,j}}\nonumber\\
&=&  \sum_{n,j}\hat{\bu}_{n,j}^{Y} \bL^*\otimes \OL{\bY_{n,j}} +\hat{\bu}_{n,j}^{\psi} \bL^*\otimes \OL{\bPsi_{n,j}} + \hat{\bu}_{n,j}^{\phi} \bL^*\otimes \OL{\bPhi_{n,j}}
=  \bL^*\otimes\OL{\bu}\nonumber
\end{eqnarray}

To prove relation (\ref{eq:FilterDivHozTensorCommute}), we need the following identities:
\begin{eqnarray}
\grad^*\bdot\bY^{(1,1)}_{n,j} &=& \bzed~,\lb{eq:DivTensorIdentity_1}\\
\grad^*\bdot\bY^{(1,2)}_{n,j} &=& \bzed~,\lb{eq:DivTensorIdentity_2}\\
\grad^*\bdot\bY^{(1,3)}_{n,j} &=& \bzed~,\lb{eq:DivTensorIdentity_3}\\
\grad^*\bdot\bY^{(2,1)}_{n,j} &=& -\sqrt{n(n+1)} ~\bY_{n,j} + \bPsi_{n,j}~,\lb{eq:DivTensorIdentity_4}\\
\grad^*\bdot\bY^{(2,2)}_{n,j} &=& -\sqrt{2} ~\bY_{n,j} + \frac{\sqrt{n(n+1)}}{\sqrt{2}}~\bPsi_{n,j}~,\lb{eq:DivTensorIdentity_5}\\
\grad^*\bdot\bY^{(2,3)}_{n,j} &=& -\frac{\sqrt{n(n+1)-2}}{\sqrt{2}}~\bPsi_{n,j}~,\lb{eq:DivTensorIdentity_6}\\
\grad^*\bdot\bY^{(3,1)}_{n,j} &=&  \bPhi_{n,j}~,\lb{eq:DivTensorIdentity_7}\\
\grad^*\bdot\bY^{(3,2)}_{n,j} &=& -\frac{\sqrt{n(n+1)-2}}{\sqrt{2}}~\bPhi_{n,j}~,\lb{eq:DivTensorIdentity_8}\\
\grad^*\bdot\bY^{(3,3)}_{n,j} &=& - \frac{\sqrt{n(n+1)}}{\sqrt{2}}~\bPhi_{n,j}~.\lb{eq:DivTensorIdentity_9}
\end{eqnarray}
Identities (\ref{eq:DivTensorIdentity_1})-(\ref{eq:DivTensorIdentity_3}) are straightforward. To derive identities (\ref{eq:DivTensorIdentity_4}) and (\ref{eq:DivTensorIdentity_7}), we use eq. (6.39) in \cite{FreedenSchreiner09}. We also use the product rule, along with eqs. (6.75)-(6.78) in \cite{FreedenSchreiner09} to derive identities (\ref{eq:DivTensorIdentity_5}) and (\ref{eq:DivTensorIdentity_9}), as well as Lemma 5.25 in that book to get identities (\ref{eq:DivTensorIdentity_6}) and (\ref{eq:DivTensorIdentity_8}). From these identities, relation (\ref{eq:FilterDivHozTensorCommute}) becomes straightforward,
\begin{eqnarray}
\OL{\grad^*\bdot\bT} 
&=&G_\bx\overrightarrow{*}\grad^*\bdot\sum_{n=0}^\infty \sum_{j=-n}^n \sum_{i,k=1}^3 \hat{T}^{Y^{(i,k)}}_{n,j}(r) ~~ \bY^{(i,k)}_{n,j}  \nonumber\\
&=&\sum_{n=0}^\infty \sum_{j=-n}^n \sum_{i,k=1}^3 \hat{T}^{Y^{(i,k)}}_{n,j}(r) ~~ G_\bx\overrightarrow{*}\left(\grad^*\bdot\bY^{(i,k)}_{n,j}\right)  ,\nonumber
\end{eqnarray}
but for every index pair $(i,k)$, we have $G_\bx\overrightarrow{*}\left(\grad^*\bdot\bY^{(i,k)}_{n,j}\right) = \widehat{G}(n)\left(\grad^*\bdot\bY^{(i,k)}_{n,j}\right)$, using the above identities along with Lemma \ref{lem:VectorConvEigen}. We finally get,
\begin{eqnarray}
\OL{\grad^*\bdot\bT} 
&=&\grad^*\bdot\sum_{n=0}^\infty \sum_{j=-n}^n \sum_{i,k=1}^3 \hat{T}^{Y^{(i,k)}}_{n,j}(r) ~~ \widehat{G}(n)\bY^{(i,k)}_{n,j}\nonumber\\
&=&\grad^*\bdot\sum_{n=0}^\infty \sum_{j=-n}^n \sum_{i,k=1}^3 \hat{T}^{Y^{(i,k)}}_{n,j}(r) ~~ G_\bx\overleftrightarrow{*}\bY^{(i,k)}_{n,j}\nonumber\\
&=&\grad^*\bdot\OL{\bT} \nonumber
\end{eqnarray}

Relation (\ref{eq:FilterCurlDotHozTensorCommute}) follows a similar proof.
\hfill $\Box$
\\
\\
REMARK: It is worth noting that relation (\ref{eq:FilterVectorDotRadialUnitVectorCommute}) implies filtering a purely tangential field yields a purely tangential field.

We can generalize the results of Proposition \ref{prop:FilterHozDerivCommute} to 3-dimensional differential operators in
the following corollary.
\begin{Cor} For a zonal kernel $G_\bx$ on $\cS$, and any scalar field $f\in C^{(2)}(\cR^3)$, vector field $\bu\in {\bf c}^{(2)}(\cR^3)$, and tensor field $\bT\in {\bf C}^{(1)}(\cR^3)$, we have 
\begin{eqnarray}
\OL{\grad f} &=& \grad \OL{f}~,\lb{eq:FilterGradScalarCommute}\\
\OL{\grad \btimes\bu} &=& \grad \btimes\OL{\bu}~,\lb{eq:FilterCurlCommute}\\
\OL{\grad \bdot\bu} &=& \grad \bdot\OL{\bu}~,\lb{eq:FilterDivVectorCommute}\\
\OL{\Delta f} &=& \Delta \OL{f}~,\lb{eq:FilterLaplacianScalarCommute}\\
\OL{\Delta \bu} &=& \Delta\OL{\bu}~,\lb{eq:FilterLaplacianVectorCommute}\\
\OL{\grad \otimes\bu} &=& \grad\otimes \OL{\bu}~,\lb{eq:FilterGradVectorCommute}\\
\OL{\grad \bdot\bT} &=& \grad \bdot\OL{\bT}~.\lb{eq:FilterDivTensorCommute}
\end{eqnarray}
\lb{cor:FilterDerivCommute}\end{Cor}
The proof is a simple consequence of our filtering operation not acting in the radial direction. Therefore, it commutes with radial derivatives. Relation (\ref{eq:FilterCurlCommute}) follows from expressing the curl operator in coordinate free form shown in eq. (\ref{eq:Curl3DCoordFree}), 
$$\grad\btimes\bu = \frac{1}{r}\grad^*\btimes\bu + \frac{1}{r}(\grad^*u_r)\btimes\hat{\Be}_r - \frac{1}{r}\partial_r\left[r\,\bu\btimes\hat{\Be}_r\right],$$
then using results (\ref{eq:FilterGradHozScalarCommute}), (\ref{eq:FilterVectorCrossRadialUnitVectorCommute}), and (\ref{eq:FilterRadialCurlVectorCommute}) of Proposition \ref{prop:FilterHozDerivCommute}. Relation (\ref{eq:FilterLaplacianVectorCommute}) follows from results (\ref{eq:FilterGradScalarCommute})-(\ref{eq:FilterDivVectorCommute}) and the identity $\Delta \bu = \grad\left(\grad\bdot\bu\right) - \grad\btimes\left(\grad\btimes\bu\right)$.

\section{Filtering without SFTs\lb{sec:FilteringWOsfts}}
Thus far, our generalized filtering of vectors and tensors on $\cS$ requires us to perform Spherical Fourier Transforms. In this section, we shall show that this is not necessary and that the filtering operation, because it commutes with differential operators, is equivalent to a convolution of the scalar fields resultant from a Helmholtz decomposition.

\subsection{Vector fields}
The generalized filtering of vectors on $\cS$ in eq. (\ref{eq:GeneralVectorConv}) requires us to perform Spherical Fourier Transforms to decompose a vector field $\bu$ into components $\bu^{(i)}$ according to eq. (\ref{eq:EdmondsVectors}). The following proposition shows that SFTs are not necessary and that filtering vectors on $\cS$ is equivalent to filtering three scalar fields after performing a Helmholtz decomposition\footnote{Of course, a Helmholtz decomposition may be accomplished by several methods, including SFTs.}.

Following Theorem 5.4 in \cite{FreedenSchreiner09} (see also \cite{Backus67,FreedenGerhards10}), for any continuously differentiable vector field $\bu$ on $\cS$, there exist unique scalar fields $u_r\in C^{(1)}(\cS)$ and $f$, $\eta \in C^{(2)}(\cS)$, satisfying $u_r = \bu\bdot\hat{\Be}_r$, and $\langle f\rangle = \langle \eta\rangle = 0$, such that 
\be
\bu = u_r\,\hat{\Be}_r + \grad^* \,f + \grad^*\btimes(\hat{\Be}_r  \,\eta).
\lb{eq:HelmholtzDecompositionVector}\ee
The second term is the irrotational component while the third is the solenoidal (or toroidal) component of the tangential vector field.

\begin{Prop} For a zonal kernel $G_\bx$ on $\cS$, such that $G\in L^p[-1,1]$ and any continuously differentiable vector $\bu\in$ {\bf l}$^q(\cS)$ with $p,q\in(1,\infty)$ and $1/p+1/q=1$, whose Helmholtz decomposition is given by eq. (\ref{eq:HelmholtzDecompositionVector}), we have
$$
\OL{\bu} = \OL{u_r}\,\hat{\Be}_r  + \grad^* \,\OL{f} + \grad^*\btimes(\hat{\Be}_r  \,\OL\eta)
$$
\lb{prop:FilterVectorByScalar}\end{Prop}
The proof follows directly from relations in Proposition \ref{prop:FilterHozDerivCommute}.

Proposition \ref{prop:FilterVectorByScalar} has practical utility when filtering on the sphere. It also establishes an important link between our framework and standard practices in numerical weather prediction and in climate spectral modeling, where scalar fields of a flow in eq. (\ref{eq:HelmholtzDecompositionVector}) are spectrally truncated\footnote{Spectral truncation is a specific choice of the filtering kernel, $G$.}. Our approach, which is constructive, makes clear the connection between filtering in physical space and filtering in spectral space. 

\subsection{Tensor fields}
Similar to filtering vectors without SFTs, we can also use Helmholtz decomposition to represent rank-2 tensors as scalar fields on which the filter operates. Following Theorem 6.6 in \cite{FreedenSchreiner09} (see also \cite{Backus66,Backus67}), for any continuously twice differentiable tensor field $\bT$ on $\cS$, there exist unique scalar fields, $F^{(i,k)}\in  C^{(2)}(\cS)$, where $i,k=1,2,3$, such that 
\begin{eqnarray}
 \bT 
&=& \left [\hat{\Be}_r\otimes\hat{\Be}_r F^{(1,1)} \right ]
+ \left [\hat{\Be}_r\otimes\grad^* F^{(1,2)} \right ]
+ \left [\hat{\Be}_r\otimes\bL^* F^{(1,3)}\right ]\nonumber\\
&+& \left [\grad^*F^{(2,1)} \otimes \hat{\Be}_r\right ]
+ \left [\bL^*F^{(3,1)}\otimes\hat{\Be}_r\right ]\nonumber\\
&+& \left [\grad^*\otimes(\hat\Be_r F^{(2,2)}) -\grad^*F^{(2,2)}\otimes\hat\Be_r \right ]\nonumber\\
&+& \left [ \grad^*\otimes\grad^*F^{(2,3)} - \bL^*\otimes\bL^*F^{(2,3)}+ 2\grad^*F^{(2,3)} \otimes\hat\Be_r \right ]\nonumber\\
&+& \left [\grad^*\otimes\bL^*F^{(3,2)} + \bL^*\otimes\grad^*F^{(3,2)}+ 2\bL^*F^{(3,2)} \otimes\hat\Be_r  \right ]\nonumber\\
&+& \left [ \bL^*\otimes(\hat\Be_r F^{(3,3)}) -\bL^* F^{(3,3)}\otimes\hat\Be_r  \right ],
\lb{eq:HelmholtzDecompTensor}\end{eqnarray}
where $\langle F^{(i,k)} Y_{0,0}\rangle=0$ for $(i,k)=(1,2),(1,3),(2,1),(2,3),(3,1),(3,2)$, and \newline
$\langle F^{(i,k)} Y_{1,j}\rangle=0$ for $j=-1,0,1$ and $(i,k)=(2,3),(3,2)$. It is worth noting that the proof of Theorem 6.6 in \cite{FreedenSchreiner09} is constructive. It solves for the scalar fields using the Green's function of the Beltrami operator and its iteration. With this decomposition, we can arrive at the following result.

\begin{Prop} For a zonal kernel $G_\bx$ on $\cS$, such that $G\in L^p[-1,1]$ and any continuously twice differentiable tensor $\bT\in$ $\bL^q(\cS)$ with $p,q\in(1,\infty)$ and $1/p+1/q=1$, whose Helmholtz decomposition is given by eq. (\ref{eq:HelmholtzDecompTensor}), we have
\begin{eqnarray}
\OL{\bT}
&=& \left [\hat{\Be}_r\otimes\hat{\Be}_r \OL{F^{(1,1)}} \right ]
+ \left [\hat{\Be}_r\otimes\grad^* \OL{F^{(1,2)}} \right ]
+ \left [\hat{\Be}_r\otimes\bL^* \OL{F^{(1,3)}}\right ]\nonumber\\
&+& \left [\grad^*\OL{F^{(2,1)}} \otimes \hat{\Be}_r\right ]
+ \left [\bL^* \OL{F^{(3,1)}}\otimes\hat{\Be}_r\right ]\nonumber\\
&+& \left [\grad^*\otimes(\hat\Be_r \OL{F^{(2,2)}}) -\grad^*\OL{F^{(2,2)}}\otimes\hat\Be_r \right ]\nonumber\\
&+& \left [ \grad^*\otimes\grad^*\OL{F^{(2,3)}} - \bL^*\otimes\bL^*\OL{F^{(2,3)}}+ 2\grad^*\OL{F^{(2,3)}} \otimes\hat\Be_r \right ]\nonumber\\
&+& \left [\grad^*\otimes\bL^*\OL{F^{(3,2)}} + \bL^*\otimes\grad^*\OL{F^{(3,2)}}+ 2\bL^*\OL{F^{(3,2)}} \otimes\hat\Be_r  \right ]\nonumber\\
&+& \left [ \bL^*\otimes(\hat\Be_r \OL{F^{(3,3)}}) -\bL^* \OL{F^{(3,3)}}\otimes\hat\Be_r  \right ],
\end{eqnarray}
\lb{prop:FilterTensorField}\end{Prop}
\vspace{.3cm}
The proof follows directly from relations in Proposition \ref{prop:FilterHozDerivCommute}.

\section{Conclusion\lb{sec:Conclusion}}
We have introduced a new definition for filtering on $\cS$ based on a generalization of the convolution of vectors and tensors. We proved that our filtering operation is linear and mean-preserving, which are two basic properties desired of a filter, and more importantly that it commutes with differential operators on the 2-sphere. Furthermore, any vectors and tensors that are normal or tangent to $\cS$ remain so after filtering. The results here hold for a generic convolution kernel that is a zonal function in $L^p[-1,1]$, without any other restrictions or assumptions. Our approach relied on the theory of spherical harmonics, which allowed us to avoid using a particular coordinate system and pole singularities. However, we also showed that the new filtering operation does not require Spherical Fourier Transforms if a Helmholtz decomposition of vectors and tensors is carried out. 

This paper lays the mathematical groundwork for developing a rigorous scale-analysis and modeling framework of PDEs on the sphere. The implementation of this framework is already underway in applications to realistic global oceanic flows \cite{Aluieetal18,Sadeketal17} to understand and quantify the \emph{dynamical} interactions between length scales.
While previous efforts have focused for the most part on the scale analysis of data, our motivation in this work stems from the analysis of \emph{scale-dynamics} on $\cS$ as described in the Introduction. The commuting property is essential to doing so for it allows the filtered equations to resemble the original PDEs, therefore, making its analysis tractable. 
The results here can also be a useful contribution to approximation theory, where there are promising and exciting efforts to use radial basis functions for solving PDEs on the sphere \cite{LeGia04,LeGia05,FlyerWright07,FlyerWright09,FlyerFornberg11,FornbergFlyer15a,FornbergFlyer15b}.

\section*{Acknowledgments}
I thank Matthew Hecht and Geoffrey Vallis for valuable discussions on oceanic flows that spurred questions leading to this paper. 

\bibliographystyle{siamplain}
\bibliography{math}
\end{document}

%% file: ex_shared.tex
\usepackage{lipsum}
\usepackage{amsfonts}
\usepackage{graphicx}
\usepackage{epstopdf}
\usepackage{algorithmic}
\usepackage{pdfpages}
\usepackage{amssymb}
\usepackage{graphicx,pdfpages,epsfig}
\usepackage{latexsym}
\usepackage{mathtools}

\newtheorem{Lem}{Lemma}
\newtheorem{Cor}{Corollary}
\newtheorem{Prop}{Proposition}
\newtheorem{Property}{Property}
     
\newcommand{\be}{\begin{equation}}
\newcommand{\ee}{\end{equation}} 
\newcommand{\lb}{\label}

\newcommand{\OL}{\overline}

\newcommand{\const}{({\rm const.})}

\newcommand{\Be}{{\bf e}}

\newcommand{\br}{{\bf r}}
\newcommand{\bu}{{\bf u}}
\newcommand{\bw}{{\bf w}}

\newcommand{\bv}{{\bf v}}
\newcommand{\bx}{{\bf x}}
\newcommand{\by}{{\bf y}}

\newcommand{\bK}{{\bf K}}
\newcommand{\bL}{ { \bf L}}

\newcommand{\bT}{{\bf T}}
\newcommand{\bY}{{\bf Y}}
\newcommand{\bPsi}{{\bf \Psi}}
\newcommand{\bPhi}{{\bf \Phi}}

\newcommand{\cR}{{\mathbb R}}
\newcommand{\cN}{{\mathbb N}_0}
\newcommand{\cS}{{{\mathbb S}^2_r}}

\newcommand{\grad}{{\mbox{\boldmath $\nabla$}}}
\newcommand{\bdot}{{\mbox{\boldmath $\cdot$}}}

\newcommand{\btimes}{{\mbox{\boldmath $\times$}}}
\newcommand{\bzed}{{\mbox{\boldmath $0$}}}

\newcommand{\TheTitle}{Convolutions on the Sphere:\\ Commutation with Differential Operators\hspace{.5cm}} 
\newcommand{\TheAuthors}{Hussein Aluie}

\headers{\TheTitle}{\TheAuthors}

\title{{\TheTitle}\thanks{This work was supported by NASA grant 80NSSC18K0772  and DOE Office of Fusion Energy Sciences grant DE-SC0014318. Partial support was also provided by an initial seed grant from the Institute of Geophysics, Planetary Physics, and Signatures (IGPPS) at Los Alamos National Laboratory (LANL), NSF Grant OCE-1259794, and DOE NNSA award DE-NA0001944.}}

\author{
  Hussein Aluie\thanks{Department of Mechanical Engineering
  \and Laboratory for Laser Energetics,
  University of Rochester, Rochester, New York 14627, USA
    (\email{hussein@rochester.edu}, \url{http://www.complexflowgroup.com}).}
}

\usepackage{amsopn}